\documentclass[final,3p,times]{elsarticle}

\usepackage[utf8]{inputenc}
\usepackage[T1]{fontenc}
\usepackage{lmodern,textcomp} 

\graphicspath{{graphics/}}

\DeclareGraphicsExtensions{.pdf,.jpeg,.png}

\usepackage{epsfig}

\usepackage{amsmath}
\usepackage{amsfonts}
\usepackage{amssymb}
\usepackage{float}

\usepackage[normalem]{ulem}

\usepackage{booktabs}
\usepackage{tabularx}
\usepackage{threeparttable}
\usepackage{siunitx}
\usepackage{graphicx}
\usepackage{subfig}

\usepackage{url}
\usepackage[colorlinks=true, citecolor=blue, linkcolor=blue, filecolor=blue,urlcolor=blue]{hyperref}

\usepackage[gen]{eurosym}

\usepackage{tikz}

\usepackage[europeanresistors,americaninductors]{circuitikz}
\usepackage{adjustbox}

\usepackage{fixltx2e}

\journal{IEW}

\begin{document}
\begin{frontmatter}



\title{Revenue and risk of variable renewable electricity investment: the cannibalization effect under high market penetration}



\author[chalmers,aalto]{L.~Reichenberg\corref{cor1}}
\ead{lina.reichenberg@chalmers.se}
\author[FMI]{T.~Ekholm}
\author[KU]{T.~Boomsma}

\cortext[cor1]{Corresponding author}
\address[chalmers]{Department of Space Earth and Environment, Chalmers University of Technology, 412 96, Göteborg, Sweden}
\address[FMI]{Finnish Meteorological Institute, Helsinki, Finland}
\address[KU]{Department of Mathematical Sciences, University of Copenhagen, Denmark}
\address[aalto]{Department of Mathematics and Systems Analysis, Aalto University, Otakaari 1 F, Espoo, Finland}

\begin{abstract}
The market revenue for variable renewable electricity (VRE) assets has been under intense scrutiny during the last few years. The observation that wind and solar power depress market prices at times when they produce the most has been termed the 'cannibalization effect'. This can have a substantial impact on the revenue of these technologies, the magnitude of which has already been established within the economic literature on current and future markets. 
Yet, the effect is neglected in the capital budgeting literature assessing green investments in the electricity sector (e.g.\ including methods such as portfolio- and real-options theory). In this paper, we present an analytical framework that explicitly models the correlation between VRE production and electricity price, as well as the impact on revenues of the surrounding capacity mix and cost to emit CO$_2$. In particular, we derive closed-form expressions for the short-term and long-term expected revenue, the variance of the revenue and the timing of investments. 
The effect of including these system characteristics is illustrated with numerical examples, using a wind investment in the Polish electricity system as a test case. We find the cannibalization effect to have major influence on the revenues, making the projected profit of a project decrease from $33$\% to between $13$\% and $-40$\% (i.e.\ a loss), depending on the assumption for the rate of future VRE capacity expansion. Using a real options framework, the investment threshold increases by between $13$\% and $67$\%, due to the inclusion of cannibalization. Our results likewise indicate that subjective beliefs and uncertainty about the future electricity capacity mix, e.g.\ VRE capacity growth, significantly affect the assessment of the revenue and investment timing. 
\end{abstract}


\end{frontmatter}

\section*{Highlights}

\begin{itemize}
  \item An analytical framework was developed to account for the effect of cannibalization in the investment decision for VRE capacity
  \item A numerical example shows that the cannibalization effect has substantial impact on expected revenues
  \item Beliefs concerning the VRE growth rate likewise has a large impact on the expected revenues and optimal timing of investment
  \item These findings may be used in various frameworks for investment decision analysis
\end{itemize}

\section{Introduction}

With stringent climate targets \cite{delbeke2015eu, european2018clean} and increasing prices on emission permits \cite{flachsland2020avoid,friedrich2020fundamentals}, the European energy sector is in rapid transition. This is especially true for the electricity sector, offering a large, cost-efficient potential for reducing greenhouse gas emissions \cite{Ekholm2010}. The electricity sector is characterized by long investment horizons; and in the case of carbon-neutral technologies, high investment costs and low running costs. For wind and solar power, the investment costs make up around 75\% of the discounted lifetime costs \cite{vartiainen2020impact}. Due to long payback periods, the financial risk for investors of exposure to low market prices is high.  

The assessment of green investments in the electricity sector often neglect the market price dynamics that result from variable renewable electricity (VRE). With a high market penetration of wind and solar power, prices are depressed during times of high VRE production, leading to value deflation of VRE assets \cite{das2020learning,mills2015strategies}. This effect has been termed 'cannibalization' \cite{cannibal2018}, and has been observed both empirically \cite{hirth2018caused} and in various models \cite{HIRTH2013218, lamont2008assessing, winkler2016market}. Likewise, 
subjective beliefs and uncertainty about the future electricity capacity mix, e.g.\ VRE capacity growth, significantly affect market prices and thereby the assessment of investment.

Investment under uncertainty \citep{luenberger1997investment} can be addressed from several angles, including portfolio theory \citep{markowitz1952portfolio} and real options analysis \citep{DixitPindyck1994}. Such methods have also been applied to VRE investments. However, these dynamics have not been recognized in the capital budgeting literature, although the cannibalization effect and the decreasing market value of variable renewables have been classified as major risks for VRE investors \cite{HIRTH2013218, lamont2008assessing, winkler2016market, brown2020decreasing}. Baringo and Conejo, for example, identify three major risks faced by an investor in wind power as 'production variability [...], the eventual future decline in wind power investment costs, and the significant financial risk' \cite{6247489}. Other studies of VRE investments consider the uncertainty in investment costs \cite{5281848, kinias2017investment}, fuel prices and demand sensitivity \cite{kumbarouglu2008real} and climate policy \cite{yang2008evaluating}. There exists recent research focusing on regulatory uncertainty regarding support schemes, such as feed-in tariffs \cite{RITZENHOFEN201676} other subsidy schemes \cite{boomsma2012renewable,boomsma2015market, kitzing2014risk} and the withdrawal of subsidies \cite{adkins2016subsidies}. A few studies include some aspects of price risk in relation to VRE investment. For example, Boomsma et al. \cite{boomsma2012renewable} use price volatility to assess the risk of a VRE investment and Fleten et al. \cite{fleten2007optimal} further account for the correlation between demand and VRE output. Nevertheless, to the best of our knowledge, the future VRE penetration level and its impact on its own revenues has been neglected in this strand of literature.

More specifically, we note that all of the cited studies model the electricity price as independent of both the VRE generation capacity and the output pattern of VRE generators; either as a stochastic process or using historical prices. The electricity price is modelled as exogenous, implicitly assuming no correlation between the price and the generation pattern of a potential investment asset. Thus, by design, the methods of references \cite{boomsma2012renewable,boomsma2015market, kitzing2014risk, adkins2016subsidies, 5281848, kinias2017investment, RITZENHOFEN201676} cannot unveil the risk pertaining to cannibalization. To capture the cannibalization effect, the methodology has to account for the relationship between VRE generation and market price, either by an endogenously generated price, or by assuming that VRE generation and price are dependent through some other mechanism. We have found one paper where the price has a negative dependence on the amount of 'green electricity' capacity in the market, namely Bigerna et al. \cite{bigerna2019green}. However, the paper does not further elaborate on the mechanism, nor how it affects the results. Similarly as for the real options studies above, a review of portfolio theory for electricity generation investment found no analyses using such endogenous price formation \cite{odeh2018portfolio}.







Here we formalize the notion of the cannibalization effect and incorporate it into an analytical framework to be used for investment analysis. Moreover, we consider revenues from the entire life time of a plant. We capture the level of the electricity price at the time of sales, not only in view of the present capacity mix or historic price levels and volatility, but based on future expectations of these quantities. In our framework, the VRE revenues depend on the fluctuating generation of the VRE asset, as well as changes in the merit order curve and the correlation between the asset’s generation and the aggregate VRE generation on the market over the lifetime of the asset.

To illustrate this new perspective, and contrast it to other approaches, we compare three cases throughout the paper: 
\begin{itemize}
    \item \textbf{Case 1}: neither the merit order nor the cannibalization effects are considered. As noted above, most of the existing literature makes this assumption, e.g.\ by using exogenous prices based on historical data from markets with low shares of VRE.
    \item \textbf{Case 2}: the merit order effect is considered, but the cannibalization effect is not. This is the case for modeling approaches with exogenous price curves based on historical data from a market with a high share of VRE.
    \item \textbf{Case 3}: both the merit order and the cannibalization effects are considered. This can be done using endogenous price formation where the momentary VRE generation affects the price, and is the novelty of our framework. 
\end{itemize}

Our aim is to describe the lifetime revenues of a new investment in a VRE asset. We derive an analytical expression for the mean and variance of the lifetime revenue and incorporate our modeling into an investment timing problem. The model is illustrated with a number of examples, using the Polish power system as a test case. The country has recently experienced a surge of VRE investments and is therefore likely to experience a stronger cannibalization effect in the future, in particular if the VRE investment boom continues.

\section{Method}
\label{sec:Problem set-up}
\subsection{Model set-up and assumptions}
\label{sec:model_setup}

Consider a wholesale, energy-only electricity market, in which the supply-side includes a mix of VRE and dispatchable generation. We assume that bidding prices are determined by the marginal costs of generation. The electricity price is then determined by the short-term market equilibrium, at the intersection of the merit-order (MO) curve and demand. The amount of electricity generated with VRE fluctuates over time. In the longer term, the generating capacities of both VRE and dispatchable generation vary over time, and the same do the generating costs of dispatchable generation, which lead to a gradually changing MO curve.\\

We make the following simplifications in our model:
\begin{itemize}
    \item Demand remains constant throughout a year and over the lifetime of the plant. 
    \item There is neither electricity storage nor trade with outside markets.
    \item The MO curve consists of two parts separated by the intersection with the first axis, which represent VRE and dispatchable generation, respectively. 
    \item The market is perfectly competitive, i.e.\ investors cannot affect the price through strategic bidding. Accordingly, we assume that bidding prices are determined by the marginal costs of generation.
\end{itemize}
The effects of these simplifications and limitations of the model are discussed later in the paper.\\

\begin{figure}[!bht]
\centering
    \includegraphics[width=0.9\linewidth,clip=true]{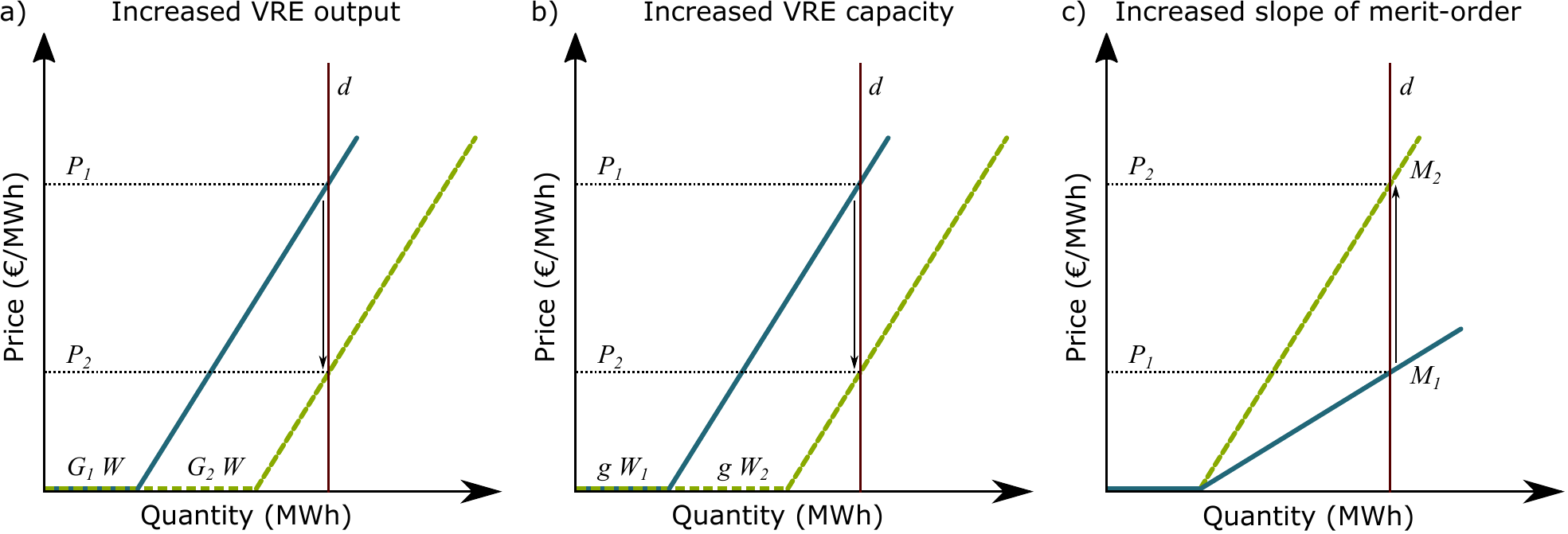}
\caption{The effects of changes in the merit-order curve on the electricity price. The fixed demand $d$ is represented by the red line. The blue lines show the initial merit-order curve leading to price $P_1$, and the green dashed lines the new merit order curve leading to price $P_2$. In a), the momentary VRE capacity factor increases from $G_1$ to $G_2$. In b), the VRE generation capacity increases from $W_1$ to $W_2$. In c), the slope of the dispatchable generators' merit order curve increases from $M_1$ to $M_2$, e.g.\ due to an increase in a CO$_2$ price.}
\label{fig:merit_order}
\end{figure}

Figure \ref{fig:merit_order} illustrates the short-term market equilibrium. The MO curve comprises two line segments that represent respectively the VRE and dispatchable generation in the market. The charts show how price is affected by the momentary changes in VRE output (left), longer term changes in VRE capacity (middle) and the steepness of the part of the MO curve representing dispatchable generation (right). The VRE output depends on the installed capacity and the momentary generating conditions (e.g. windiness or solar irradiance). The steepness of the dispatchable generation MO curve depends on the aggregate generating capacity and its mix, and on fuel and emission prices as well as other variable operating costs. As can be seen from the figure, the equilibrium price decreases with an increase in VRE output and capacity level, whereas the price increases with an increase in the costs of dispatchable generation.

Starting from the static description illustrated in Figure \ref{fig:merit_order}, we develop an analytical model that computes the expected value and variance of the net present value (NPV) of future revenues for a unit of VRE capacity.
The model encapsulates that the aggregate installed generation capacity mix can change over time due to new investments and decommissioning of old plants, and that the generation costs of dispatchable generation change due to fluctuations in fuel and emission prices. 
We model these as stochastic processes that affect the steepness of the MO curve and the intersection with the first axis, representing the costs of dispatchable generation and level of VRE capacity, respectively. Furthermore, we include a stationary stochastic processes that represents the momentary variability in VRE generation.

We derive an expression for the expected revenues in two steps: First we consider the revenue at a single future point in time, see Section \ref{sec:Short-term revenue}. This static approach reflects the instantaneous uncertainty in VRE generation, while treating the capacity mix (determined by the VRE capacity and dispatchable MO curve) as known quantities. Then we consider long-term changes in VRE capacity and the MO curve in Section \ref{sec:Long-term revenue}, and consider the expected net present value (NPV) of revenues over the lifetime of the VRE asset. We likewise derive an expression for the variance of the revenue NPV over the lifetime. We apply the analytical framework to an investment timing problem in Section \ref{sec:investment_timing}. All three components of our modeling (instantaneous revenue, life-time revenue and risk, and investment timing) are illustrated with a numerical example, using real-world data for Poland.



\subsection{Nomenclature}

The symbols used in the analytical model are presented in \ref{tbl:nomenclature} for easy reference.

\begin{table}[hbt!]
\caption{Nomenclature with descriptions and units. For the definition of constants $p_1$ to $p_6$, please see \ref{sec:AppVariance}.}
\label{tbl:nomenclature}
\begin{center}
 \begin{tabular}{l l l} 
 \hline
   Symbol  & Description & Unit  \\ 
   \hline
 $t$ & time  & time unit e.g. [h] \\ 
 $T$ & lifetime of investor's plant & time unit e.g. [h] \\ 
 $V_t$ & Wind power output time series  & [0,1] \\ 
 $S_t$ & Solar power output time series  & [0,1] \\
 $G_{A_t}$ & Aggregated VRE output & [0,1] \\
 $G_{I_t}$ & Investor VRE output & [0,1] \\
 $\rho_{G_A, G_I}$ & Pearson correlation coefficient between $G_{A_t}, G_{I_t}$  & [-1,1] \\
 $W_t$, $w_t$ & total VRE capacity & [GW] \\
 $W_t$, $w_t$ & total VRE capacity & [GW] \\
 $M_t$, $m_t$ & slope of merit order curve & [€/MWh/GW] \\
 $d$ & electricity demand & [GW] \\
  $P_t$ & electricity price & [€/MWh] \\
 $R_t$ & revenue for VRE investor & [€/MW/h] \\
 $\mu_{G_I}$, $\mu_{G_A}$ & expected value for $G_I$ and $G_A$ & [0,1] \\
 $\sigma_{G_I}$, $\sigma_{G_A}$ & standard deviation for $G_I$ and $G_A$ & [0,1] \\
 $k_1$ & constant, $d \mu_{G_I}$ & [GW] \\
 $k_2$ & constant, $\mu_{G_A} \mu_{G_I}$ & [0,1] \\
 $k_3$ & constant, $\rho_{G_A G_I} \sigma_{G_I} \sigma_{G_A}$ & [-1,1] \\
 $\mu_W, \mu_M$ & relative drifts of stochastic variables $M_t$, $W_t$ & [-$\infty$, $\infty$] \\
 $\sigma_W, \sigma_M$ & relative volatilities of stochastic variables $M_t$, $W_t$ & [-$\infty$, $\infty$] \\
 ${z_{W,t}}$, ${z_{M,t}}$ & Wiener processes for $M_t$, $W_t$ & time unit e.g. [h]\\
$\rho_{W M}$ & correlation coefficient between $W_t$ and $W_m$, $E[dz_{W,t} dz_{M,t}]/dt$ & [-1,1]\\
$\beta$ & Discount rate & [0,$\infty$]\\
 \hline
\end{tabular}
\end{center}

\end{table}
\clearpage
\subsection{Numerical example and data}
\label{sec:Numerical_example_setup}

The analytical model is illustrated with numerical examples based on weather, demand and capacity mix data for Poland. These are used for all the numerical examples (Sections \ref{numerical_example_instantaneous revenue},\ref{Numerical_example_longterm} and \ref{Numerical_example_investmenttiming}). In addition, an estimate regarding the \emph{change} in VRE capacity and slope of the merit order curve, used in the examples in Sections \ref{Numerical_example_longterm} and \ref{Numerical_example_investmenttiming}, is based on data on the development in Germany between the years 2005 and 2019.

The statistical properties (see Table \ref{tbl:statistics_VRE}) of the time series for wind output, $V_{t} \in [0,1]$, and solar output, $S_{t} \in [0,1]$, are constructed using ERA 5 weather data from ECMWF and the DTU Global Wind Atlas with the method described in Mattsson et al. \cite{mattsson2020autopilot}. Sites in Poland with a capacity factor of solar above 14\% or average wind speed above $6$ m/s \footnote{Referring to the partitioning into classes in reference \cite{mattsson2020autopilot}, this is equivalent to averaging over classes $3-5$ for wind and class $2$, which is the highest class in Poland, for solar.} were aggregated to represent the respective wind and solar output on country level. 

To represent the statistical properties (see Table \ref{tbl:statistics_VRE}) of aggregated VRE output $G_{A_t}$, the wind and solar outputs are weighted using the wind capacity ($5.8$ GW \cite{Eurostat_REcapacity}) and solar capacity ($0.6$ GW \cite{Eurostat_REcapacity}) in Poland in 2018, so that $G_{A_t} = 0.91 V_t + 0.09 S_t$. The numerical examples assume the potential investment to be in wind power, so the the investor's time series is $G_{I_t} = v_t$. The statistical properties for wind, solar and the aggregate VRE generation of our data set are listed in Table \ref{tbl:statistics_VRE}. For the 91\% wind, 9\% solar VRE mix, the correlation coefficient between the aggregate VRE mix and the considered wind power asset, $\rho_{G_A, G_I}$, is $0.995$.

\begin{table}[!hbp]
\caption{Statistical properties of the wind, solar and aggregate VRE capacity factors used in the numerical examples throughout the paper. The values for wind and solar are based on the wind/solar share in Poland in 2018, with a VRE capacity mix of 9\% solar and 91\% wind.}
\label{tbl:statistics_VRE}
\begin{center}
 \begin{tabular}{l c c} 
 \hline
  & Mean ($\mu$) & Standard deviation ($\sigma$)  \\ 
 \hline
 Wind, $v_t$ & 0.31  & 0.22 \\ 
 \hline
 Solar, $s_t$ & 0.15  &  0.22 \\
 \hline
 Aggregate VRE, $G_{A_t}$ & 0.30 & 0.20 \\
 \hline
\end{tabular}
\end{center}

\end{table}

The initial VRE capacity $w_0$ is the sum of wind and solar capacities ($5.8 + 0.6 = 6.4$ GW in 2018 \cite{Eurostat_REcapacity}). The demand $d$ is estimated by dividing the annual demand in Poland in 2018 by the number of hours, equaling $18.5$ GW.

\begin{figure}[!tbp]
  \centering
  \includegraphics[width=0.7\textwidth]{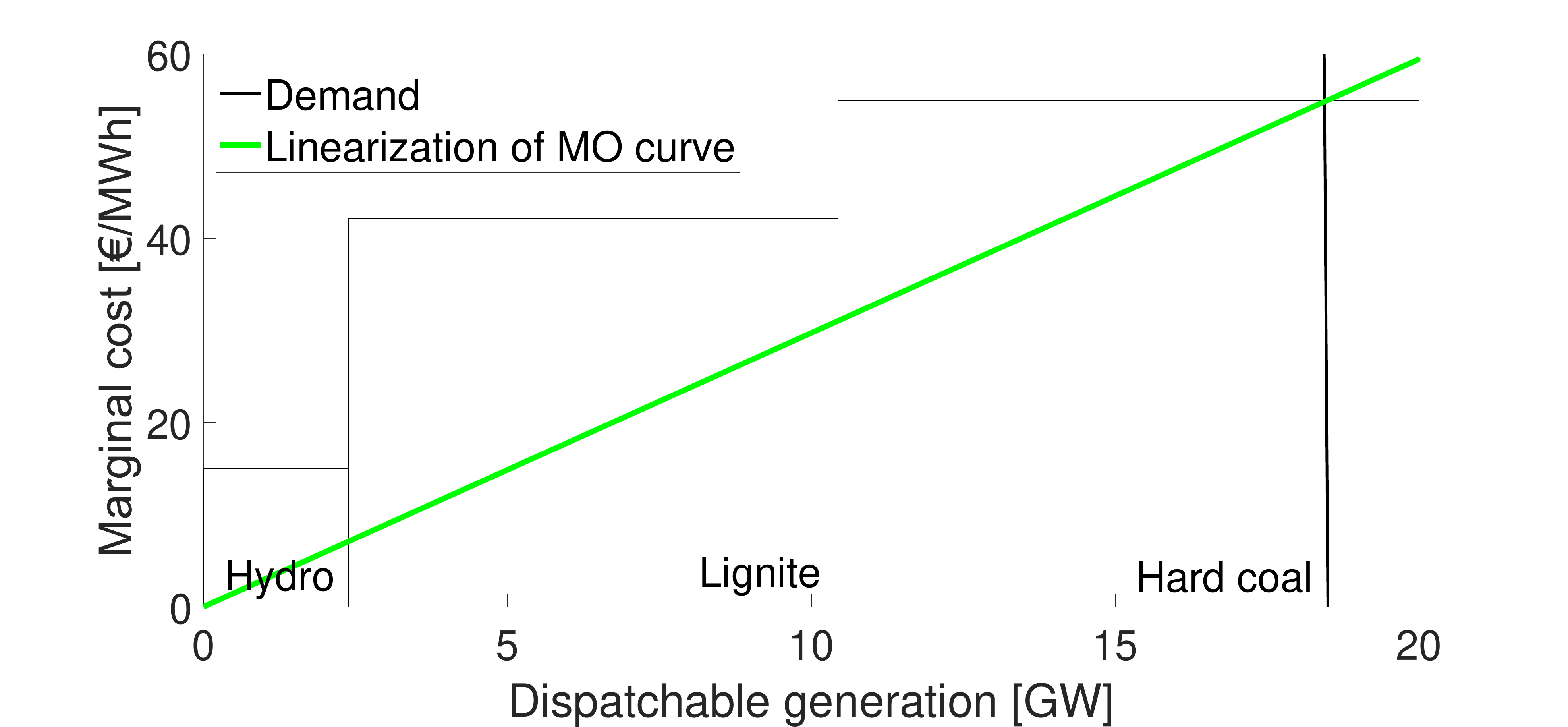}\label{fig:f1}
  \caption{The linearization of the dispatchable generation merit-order curve for Poland in 2018. The linearization is done between the origin and the level of average demand $d$.}
\label{fig:illustration_linear}
\end{figure}

The initial slope $m_0$ of the MO curve representing dispatchable generation is based on a bottom-up estimate of the generation mix in Poland. The capacities of hydro power are from \cite{Eurostat_REcapacity} and all other capacities from \cite{factsheet2019available}. The variable cost pertaining to each dispatchable technology is found by adding fuel-, variable O\&M- and CO$_2$ cost, see \ref{Appdata} for details and data sources.
We finally determine the slope $m_0$ by linearizing the resulting dispatchable MO curve between the origin and dispatchable generation at level $d$ (18.5~GW), where hard coal is at the margin with the variable cost of $55$~€/MWh (see Figure \ref{fig:illustration_linear}). The slope is found to be $m_0 = 3\cdot10^{-3}$  $\frac{\text{€}/MWh}{\text{GW}}$.


The long-term evolution of VRE and dispatchable generation capacity in Sections \ref{Numerical_example_longterm} and \ref{Numerical_example_investmenttiming} is represented by random walks using Geometric Brownian Motions. The growth and volatility of these random walks reflect the investor's subjective beliefs about the future, and definite values cannot therefore be assigned. Instead, we illustrate our results for a range of these parameters. We draw an analogy from the past: the rapid expansion of VRE capacity in Germany; with an underlying idea that Poland might be in the early stage of similar progress. The share of VRE generation of total electricity demand in Poland 2018, approximately 10\% of the annual electricity generation, is equal to that of Germany in 2005.

The average, annual relative change in VRE capacity is found to be 13\% in Germany between the years 2005 and 2019, with a standard deviation of 6\%. However, there was a significant, declining drift in the VRE expansion rate, with years around 2010 experiencing over 20\% growth, whereas years towards 2019 had growth well below 10\%. The dispatchble MO curve slope grows annually on average by 1.1\%, having a standard deviation of 4.8\%. The correlation between these is -14\%. 






\section{Instantaneous revenue}
\label{sec:Short-term revenue}

We start by considering revenue at a single future point in time.

\subsection{Derivation}


Consider the market equilibrium at time $t$. In order to determine the market's equilibrium price $P_t$, let electricity demand be $d$. Let $w_t$ be the aggregate VRE capacity with the instantaneous capacity factor represented by a random variable $G_{A,t} \in [0,1]$\footnote{In our notation, all capital letters denote random variables, whereas deterministic parameters are denoted by lowercase letters.}. Then, $w_t G_{A,t}$ determines the electricity generated by VRE. For $w_t G_{A,t} \leq d$, $d - w_tG_{A,t}$ presents the dispatch of the non-VRE plants (here-after labeled 'dispatchable') according to the MO curve. Assuming that the variable generation costs of VRE are zero and the dispatchable MO curve is linear with slope $m_t>0$, the short-term equilibrium price $P_t$ described by the function $P_t = \max\{m_t (d - w_t G_{A,t}),0\}$ as illustrated in Figure \ref{fig:merit_order}. For simplicity, however, we assume the MO curve is the function:
\begin{equation}
\label{eq:p_function_g}
    P_t = m_t (d - w_t G_{A,t}),
\end{equation}
This function is linear in the non-VRE power dispatch and affine in VRE output. As a result, $P_t$ is a random variable and $G_{A,t}$ and $P_t$ are perfectly correlated. 

Let $G_{I,t} \in [0,1]$ be a random variable that represents the capacity factor of the \emph{investor's} VRE plant at time $t$, with expected value $\mu_{G_{I}}$ and variance $\sigma_{G_{I}}^2$. Similarly for the aggregate VRE capacity factor $G_{A,t}$, let the expected value be $\mu_{G_{A}}$ and variance $\sigma_{G_{A}}^2$. For simplicity, we assume that neither their means nor their variances vary over time. Note that $G_{I,t}$ and $P_t$ may not be perfectly correlated, but are indeed correlated if $G_{I,t}$ and $G_{A,t}$ are correlated. As an example, if the aggregated VRE generation in the market is dominated by wind, and if the investor is also considering a wind power investment, then $G_{A,t}$ and $G_{I,t}$ will be highly correlated. However, for solar power investment $G_{A,t}$ and $G_{I,t}$ would not be correlated. 

The revenue per unit capacity of the investor's VRE plant at time $t$ is $R_t = P_t G_{I,t}$. The expected value of the random variable $R_t$ is  
\begin{align}
    E[R_t] &= E[P_t G_{I,t}] = E[m_t (d - G_{A,t} w_t) G_{I,t}]\\
    &=  m_t (d E[G_{I,t}] -  w_t E[G_{I,t} G_{A,t}])\\ 
    &= m_t (d E[G_{I,t}] -  w_t(Cov(G_{I,t}, G_{A,t})+E[G_{I,t}] E[G_{A,t}])\\ 
    &= m_t d \mu_{G_I} - m_t w_t \mu_{G_I} \mu_{G_A} - m_t w_t\rho_{G_I, G_A} \sigma_{G_I} \sigma_{G_A}
    \label{eq:VRE_exp_revenue_shortterm}
\end{align}
where $\rho_{G_I, G_A}$ is the correlation coefficient between $G_{I,t}$ and $G_{A,t}$.


Equation \eqref{eq:VRE_exp_revenue_shortterm} shows that, with an affine merit order curve, the expected revenue for a VRE owner is affine in the amount of VRE capacity, $w_t$, in the system. Moreover, for $\rho_{G_I,G_A}\geq 0$, it is non-increasing. The rate of decrease depends on the mean ($\mu_{G_A}$) and variance ($\sigma_{G_A}$) of the aggregated VRE output, the mean ($\mu_{G_I}$) and variance ($\sigma_{G_I}$) of investor's output, as well as the correlation between aggregated VRE output and investor's output ($\rho_{G_I, G_A}$).

In Equation \eqref{eq:VRE_exp_revenue_shortterm}, the parameters $w_t$ and $m_t$ are subject to long-term change and uncertainty, whereas the remaining parameters do not change over time \footnote{These can be estimated from data on wind or solar time series. If the relative shares of wind and solar change over time, the values pertaining to the aggregated VRE output, $\mu_{G_A}, \sigma_{G_A}, \rho_{G_A,G_I}$  will change with time, but here we assume a fixed ratio between wind and solar, which entails that wind and solar have the same growth rate.}. We can thus simplify Equation \eqref{eq:VRE_exp_revenue_shortterm} to:
\begin{align}
    E[R_t] = k_1 m_t - (k_2+k_3) m_t w_t
    \label{eq:VRE_exp_revenue_shorttermsimplified}
\end{align}
where $k_{1} = d \mu_{G_I}$, $k_{2} = \mu_{G_A} \mu_{G_I}$ and $k_{3} = { \rho_{G_I, G_A} \sigma_{G_I} \sigma_{G_A}}$. Note that the revenue decreases linearly with $k_2$ and $k_3$. Thus, both the merit order- and cannibalization effects reduce expected revenue,

In order to further understand the significance of Equations \eqref{eq:VRE_exp_revenue_shortterm} and \eqref{eq:VRE_exp_revenue_shorttermsimplified} in relation to the merit order and cannibalization effects, we can separate it into three terms:

\begin{itemize}
    \item The first part, $k_1 m_t$, corresponds to the revenue for a VRE plant if there is no VRE in the system, i.e.\ if $w_t=0$, or if the decrease in revenue due to the merit-order and cannibalization effects is ignored. Considering only this term, with $k_2=k_3=0$, corresponds to our \textbf{Case 1}.
    \item The second term, $- k_2 m_t w_t$, stems from the merit order effect. This effect increases as the energy contribution from VRE, $w_t$, increases. Considering this term in addition to the first part, i.e. $k_2>0, k_3=0$, corresponds to our \textbf{Case 2}\footnote{To assume $k_3 = 0$ is justified if price and generation is uncorrelated, i.e., $\rho_{G_I, G_A}=0$. In reality, this may happen e.g.\ if the generator is geothermal electricity, which has a flat output profile. It may also be the case that the correlation is negative, $\rho_{G_I, G_A}<0$, which may happen e.g.\ if the investor's generator is solar PV in a system dominated by wind power, since solar PV has  seasonal pattern opposite that of wind.}. 
    \item The third term, $- k_3 m_t w_t$, represents the cannibalization effect. This term takes into account the correlation, $\rho_{G_I, G_A}$, between the aggregated VRE generation ($G_{A,t}$) and the generation of the investor's plant ($G_{I,t}$). If the revenue is assessed as the sum of all three terms, then the revenue is determined by the amount of VRE in the system (ME effect), as well as the timing of investors' generation compared to the aggregate (cannibalization). Including the second and third terms ($k_2>0, k_3>0$) in the revenue assessment corresponds to our \textbf{Case 3}. 
\end{itemize}

\subsection{Numerical example}
\label{numerical_example_instantaneous revenue}

The numerical example uses data from Section \ref{sec:Numerical_example_setup}.

We use the values for a prospective wind power investment in Poland from Table \ref{tbl:statistics_VRE}. The average demand in Poland is $d= 18.5$ GW. These values generate parameters $k_{1} = 5.75$ GW, $k_{2} = 0.092$ and $k_{3} = 0.044$.  Furthermore, in this example, we assume that the VRE capacity, which in 2018 was about one third of the average demand, has increased so that it is equal to the demand, $w_t=d$. The slope of the merit order curve is $m_t = 0.003$ € \footnote{Computed assuming coal on the margin, with variable cost, including the EU-ETS price, of $55$ €/MWh.}. 

Given these numbers, the average revenues are presented in Table \ref{tbl:instaneneous_revenue}. The table presents the average hourly revenue per unit investment (in €/MW/h) for the three cases, as well as the average return per unit of electricity generated (in €/MWh), using the average capacity factor $\mu_{G_I} = 0.31$. The average return per unit of electricity may be compared against the LCOE (Levelized Cost of Electricity) of wind power, which in this case is 38 €$/MWh$, using the cost data of Table \ref{tbl:LCOE_data} in \ref{Appdata}. Thus, for an investor who does no take into account the MO- and cannibalization effects and hence makes her investment assessment according to \textbf{Case 1}, it looks as though she will make an average profit of $55-38=17$ € for each MWh generated (or a 45\% profit margin). However, if the MO- and cannibalization effects are taken into account (\textbf{Case 3}), the investment calculation yields a loss of $38-31=7$ € for every MWh generated (or a negative profit of 18\%).

\begin{table}[!htbp]
\caption{Average, instantaneous revenue estimates in the three cases, calculated as per-MW-per-hour and per generated electricity.}
\label{tbl:instaneneous_revenue}
\begin{center}
 \begin{tabular}{l c c} 
 \hline
  & Revenue (per MW, per hour) & Revenue per generated MWh  \\ 
 \hline
 Case 1 & 17 € & 55 € \\ 
 \hline
 Case 2 & 12 € & 39 € \\
 \hline
 Case 3 & 10 € & 31 € \\
 \hline
\end{tabular}
\end{center}
\end{table}



\section{Long-term expected revenue and risk}
\label{sec:Long-term revenue}

In computing the short-term revenue, the level of VRE capacity in the system and the slope of the merit-order curve, $w_t$ and $m_t$, were known constants. However, these two parameters change over the lifetime of a potential investment: the VRE capacity will change as other investors decide to install new or retire existing capacity; while the slope of the merit order curve may change due to investments and decommissioning of dispatchable power plants, or due to changes in fuel and CO$_2$ prices. All of these involve notable uncertainties. In the long term, the values $m_t$ and $w_t$ may be seen as particular realizations of the random variables $M_t$ and $W_t$. 

\subsection{Derivation of the expected net-present-value of the revenue}

We represent the long-term uncertainties regarding VRE capacity and MO slope by Geometric Brownian Motions. Let $\{W_t\}$ and $\{M_t\}$ be the two stochastic processes. Then,
\begin{eqnarray}
dW_t=\mu_WWdt+\sigma_WWdz_{W,t},\\
dM_t=\mu_MMdt+\sigma_MMdz_{M,t},
\end{eqnarray}
where $\mu_W$ and $\mu_M$ are relative drifts, $\sigma_W$ and $\sigma_M$ are relative volatilities, and $\{z_{W,t}\}$ and $\{z_{M,t}\}$ are Wiener processes with $E[dz_{W,t}dz_{M,t}]=\rho_{WM}dt$. We denote the initial state at $t=0$ with $W_0 = w_0$ and $M_0 = m_0$. 

As a consequence, the level of VRE capacity $w_t$ and the slope of the merit-order curve $m_t$ are strictly positive. The Geometric Brownian motion gives rise to a lognormal distribution with parameters that change with time. \footnote{Note that the log-normal distribution means that there is a non-zero probability of values of $W_t$ that are much higher than demand and, as a consequence of our MO curve (Equation \eqref{eq:p_function_g}), produce negative prices. The validity of the model is constrained to parameterizations for which this probability remains negligible.}

Let $\{G_{A,t}\}$ and $\{G_{I,t}\}$ be stationary stochastic processes of capacity factors, i.e.\ means and variances are constant over time, as above. We assume that $G_{A,t},G_{A,s},G_{I,t}$ and $G_{I,s}$ are mutually independent for $t \neq s$.
We also assume that both $G_t$ and $G_{A,t}$ are independent of $W_t$ and $M_t$. However, $W_t$ and $M_t$ are not necessarily independent of each other.

Future revenues are discounted with a rate of $\beta$. 
By using the expression in Equation \eqref{eq:VRE_exp_revenue_shorttermsimplified} for the instantaneous expected revenues, we can derive the expected net present value of revenues over the lifetime $T$ of a unit of VRE capacity. For known values of the VRE capacity level and slope of the MO curve at time $0$, $W_0=w_0$ and $M_0=m_0$, this is:

\begin{align}
    \label{eq:VRE_exp_revenue}
    V(w_0,m_0) &= E_{G,W,M}^{w_0,m_0}\left[\int_0^T e^{-\beta t} R(G_t,W_t,M_t) dt\right]\\
    &= \int_0^T e^{-\beta t} E_{G,W,M}^{w_0,m_0}\left[ R(G_t,W_t,M_t)\right] dt\nonumber  \\
    &= \int_0^T e^{-\beta t}E_{W,M}^{w_0,m_0}\left[E_{G_t\mid W,M}^{w_t,m_t}\left[ R(G_t,w_t,m_t) \right]\right]dt\nonumber\\
    &= \int_0^T e^{-\beta t}E_{W,M}^{w_0,m_0}\left[E_{G_t}\left[ R(G_t,W_t,M_t) \right]\right]dt\nonumber\\
    &= \int_0^T e^{-\beta t} \left( k_{1}E_{M}^{m_0}\left[M_t \right]  - (k_{2}+k_{3}) E_{W,M}^{w_0,m_0}\left[ W_tM_t \right] \right) dt \nonumber \\
    &= \int_0^T e^{-\beta t} \left( k_{1}m_0 e^{\mu_M t}  - (k_{2}+k_{3}) w_0 m_0 e^{\mu_{WM} t}   \right) dt \nonumber \\
    &= m_0 \Big( k_1\frac{1-e^{-(\beta-\mu_M) T}}{\beta-\mu_M} - w_0 (k_{2}+k_{3}) \frac{1-e^{-(\beta-\mu_{WM}) T}}{\beta-\mu_{WM}}   \Big) \nonumber    
\end{align}
where we use the notation $S=\{S_t\}$ and $E_S^{s_0}[g(S_t)]=E_S[g(S_t)\mid S_0=s_0]$. 
The second equality is by Fubini's theorem, the third equality uses the law of total expectations and the independence of short-term uncertainties ($G_t$ and $G_s$ are independent for $t\neq s$), the fourth exploits the independence of short-term and long-term uncertainties ($G_t$ and $W_s$ as well as $G_t$ and $M_s$ are independent for all $t,s$) and the fifth that $\{M_tW_t\}$ is a GBM with drift $\mu_{WM}=\mu_M+\mu_W+\rho_{WM}\sigma_W\sigma_M$. Moreover, with $S$ following a GBM with drift $\mu$, the expected value of $S_t$ is $E_S^{s_0}[S_t]=s_0 e^{\mu t}$.





Expression \eqref{eq:VRE_exp_revenue} is structurally similar to Equation \eqref{eq:VRE_exp_revenue_shortterm}, and shows that expected net present value of revenue decreases linearly with $k_2$ and $k_3$. Thus, both the merit order- and cannibalization effects reduce expected revenue, in accordance with the intuition and empirical knowledge about these effects. Furthermore, if $k_2=k_3=0$ (corresponding to \textbf{Case 1}), the expected net present value of the VRE plant increases with the drift of the slope of the merit order line, $\mu_M$. A higher value for $\mu_M$ means a faster increase of the steepness of the merit order curve. If, on the other hand, $k_2+k_3>0$ (\textbf{Cases 2 and 3}), the effects of merit order and/or cannibalization are present, which reduces the increase. Furthermore, if $k_2=k_3=0$, the value of the plant is clearly unaffected by VRE capacity, whereas if $k_2+k_3>0$, the plant value decreases with the drift of the VRE capacity, $\mu_W$, i.e. a faster growth rate in the total VRE capacity in the system depresses the revenues more. For numerical investigations of expected revenues as a function of different drift coefficients $\mu_W$ and $\mu_M$, and the impacts of the merit order and cannibalization effects (setting $k_2+k_3>0$), see Section \ref{Numerical_example_longterm}.

\subsection{Derivation of the variance of the revenue}
\label{sec:Variance of revenue}

An investor is interested not only in the expected revenue, but also in risk. Here, we measure risk through the variance of discounted revenues, and develop an analytical expression for this. The variance is given by
\begin{align}
    Var(w_0,m_0) &= E_{G,W,M}^{w_0,m_0}\left[ \left( \int_0^T e^{-\beta t} R(G_t,W_t,M_t) dt \right)^2 \right] - E_{G,W,M}^{w_0,m_0}\left[\int_0^T e^{-\beta t} R(G_t,W_t,M_t) dt\right]^2
\end{align}

A closed-form expression for the variance is derived in \ref{sec:AppVariance}, arriving at:
\begin{align}
    \label{eq:VRE_var_revenue}
    Var(w_0,m_0) = & 2 m_0^2 \Bigg( 
    \begin{aligned}[t] 
                 &k_1^2  \,\, \frac{p_2 + p_1 e^{(p_1+p_2) T} - (p_1+p_2) e^{p_1 T}}{p_1 p_2(p_1 + p_2)} \\ 
    &- w_0k_1 (k_2+k_3) \,\, \frac{p_4 + p_1 e^{(p_1+p_4) T} - (p_1+p_4) e^{p_1 T}}{p_1 p_4(p_1 + p_4)} \\ 
    &- w_0k_1 (k_2+k_3) \,\, \frac{p_5 + p_3 e^{(p_3+p_5) T} - (p_3+p_5) e^{p_3 T}}{p_3 p_5(p_3 + p_5)} \\ 
    &+ w_0^2(k_2+k_3)^2 \,\, \frac{p_6 + p_3 e^{(p_3+p_6) T} - (p_3+p_6) e^{p_3 T}}{p_3 p_6(p_3 + p_6)} 
    \Bigg) 
    \end{aligned}
    \\ \nonumber &
    - m_0^2 \left( k_1\frac{1-e^{-p_1T}}{p_1} - w_0(k_2+k_3)\frac{1-e^{-p_3 T}}{p_3} \right)^2 .
\end{align}
This expression uses the following shorthands for the parameters of the GBSs and $\beta$:
\begin{align}
    &  p_1 = \mu_M - \beta                                 \\
    &  p_2 = \mu_M - \sigma_M^2 - \beta                        \nonumber \\
    &  p_3 = \mu_{WM} - \beta                           \nonumber \\
    &  p_4 = \mu_{WM} + \rho_{WM} \sigma_W \sigma_M + \sigma_M^2 - \beta                   \nonumber \\
    &  p_5 = \mu_M + \rho_{WM} \sigma_W \sigma_M + \sigma_M^2 - \beta     \nonumber \\
    &  p_6 = \mu_{WM} + \sigma_{WM}^2 - \beta     \nonumber .
\end{align}

This expression is significantly more complicated than Equation \eqref{eq:VRE_exp_revenue_shorttermsimplified} or \eqref{eq:VRE_exp_revenue}. To obtain further insights regarding the dependence of the variance on merit-order and cannibalization effects, as well as on the parameters of the GBMs, we carry out numerical investigations in the next section. 

\subsection{Numerical example}
\label{Numerical_example_longterm}

We assess the expected revenue and risk by implementing Equation \eqref{eq:VRE_exp_revenue} with parameter values for Poland, as described in Section \ref{sec:Numerical_example_setup}. In addition to the parameter values pertaining to demand, wind- and solar output, which yield the constants $k_1=5.75$ GW, $k_2=9.2*10^{-2}$, $k_3=4.4*10^{-2}$ (see Section \ref{sec:Numerical_example_setup}), we assume the starting values for the Polish system in 2018 to be $M_0=3*10^-3$ and $W_0=6.4$ GW, again in accordance with the assessment in Section \ref{sec:Numerical_example_setup}. We set the discount rate, $\beta$, to $5$\%. 

We estimate plausible values for the (future) relative drifts ($\mu_W$ and $\mu_M$) and relative volatilities ($\sigma_W$ and $\sigma_M$) of the VRE capacity and the merit-order slope on the basis of the German electricity system between the years 2005 and 2019, as presented in Section \ref{sec:Numerical_example_setup}. An exception is the average VRE growth rate $\mu_W$. Germany experienced an unprecedented VRE capacity expansion during this period, on average 13\% per year; fuelled partly by strong subsidy schemes, and which has lead e.g. to more frequent negative prices \citep[see e.g.][]{Khoshrou2019}.
Negative prices can also become a problem in terms of the validity of our model, and this starts to become significant around 10\% average VRE growth rate (see \ref{sec:validation}). 
Therefore, we use a more conservative assumption of 5\% average VRE capacity expansion as the default case and 10\% as the high-end of the range. We calculate the expected revenues and their standard deviation by varying one of the random walk parameters at a time, while keeping the others at their default values. The parameter ranges and the default values are presented in Table \ref{tbl:GBM_parameters}. 

\begin{table}[!hbp]
\caption{The parameter ranges and default values for the Geometric Brownian Motions for $\{W_t\}$ and $\{M_t\}$.}
\label{tbl:GBM_parameters}
\begin{center}
 \begin{tabular}{l c c c} 
 \hline
  & Default & Range min & Range max  \\ 
 \hline
 VRE growth $\mu_W$             & 5 \%  & 0 \% & 10 \% \\ 
 VRE volatility $\sigma_W$      & 6 \%  & 0 \% & 10 \% \\ 
 MO slope growth $\mu_M$        & 1 \%  & -5 \% & 5 \% \\ 
 MO slope volatility $\sigma_M$ & 5 \%  & 0 \% & 10 \% \\ 
 VRE and MO correlation $\rho_{WM}$ & -10 \% & -- & -- \\ 
 \hline
\end{tabular}
\end{center}
\end{table}

\begin{figure}[!hbp]
\centering
    \includegraphics[width=0.8\linewidth]{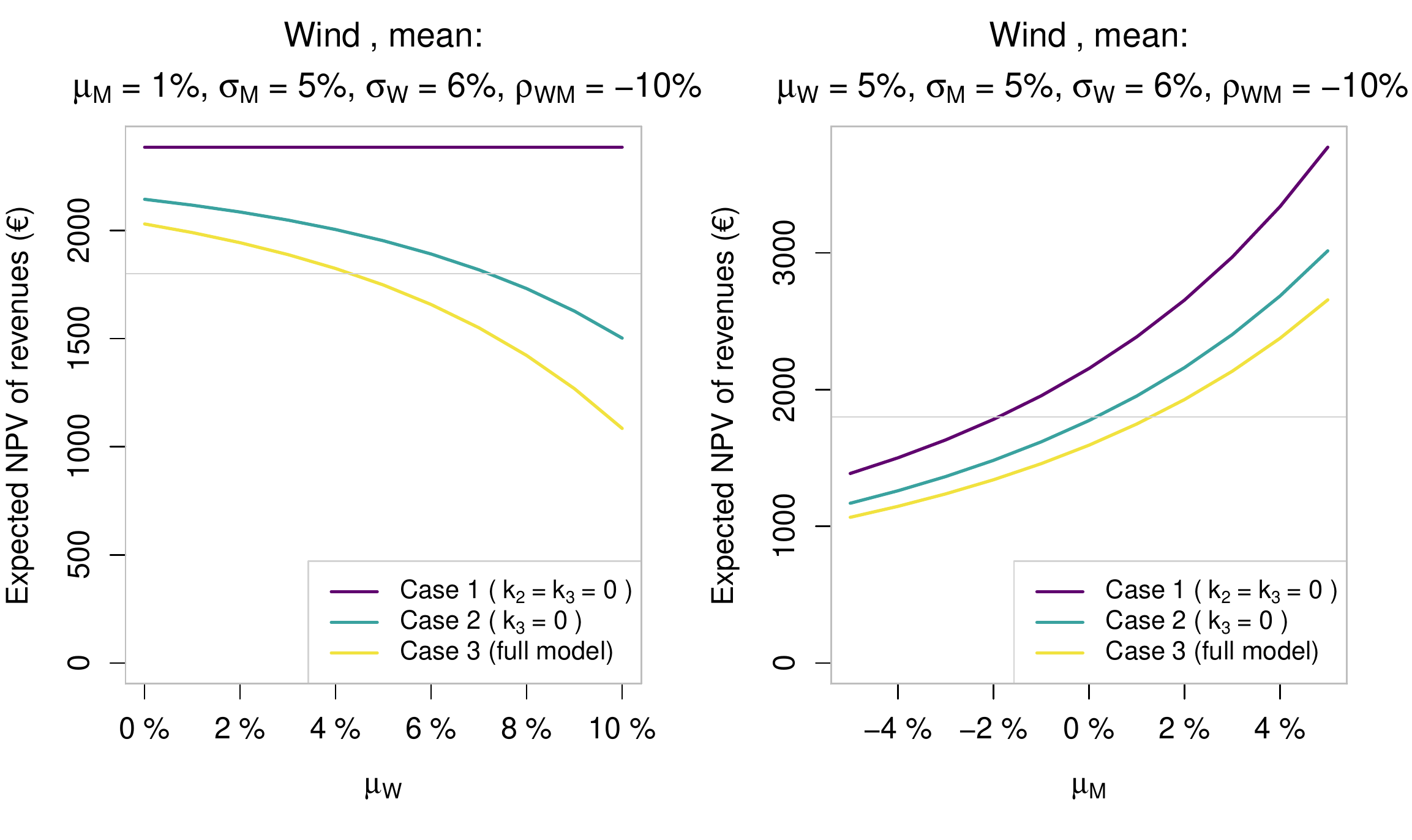}
    \includegraphics[width=0.8\linewidth]{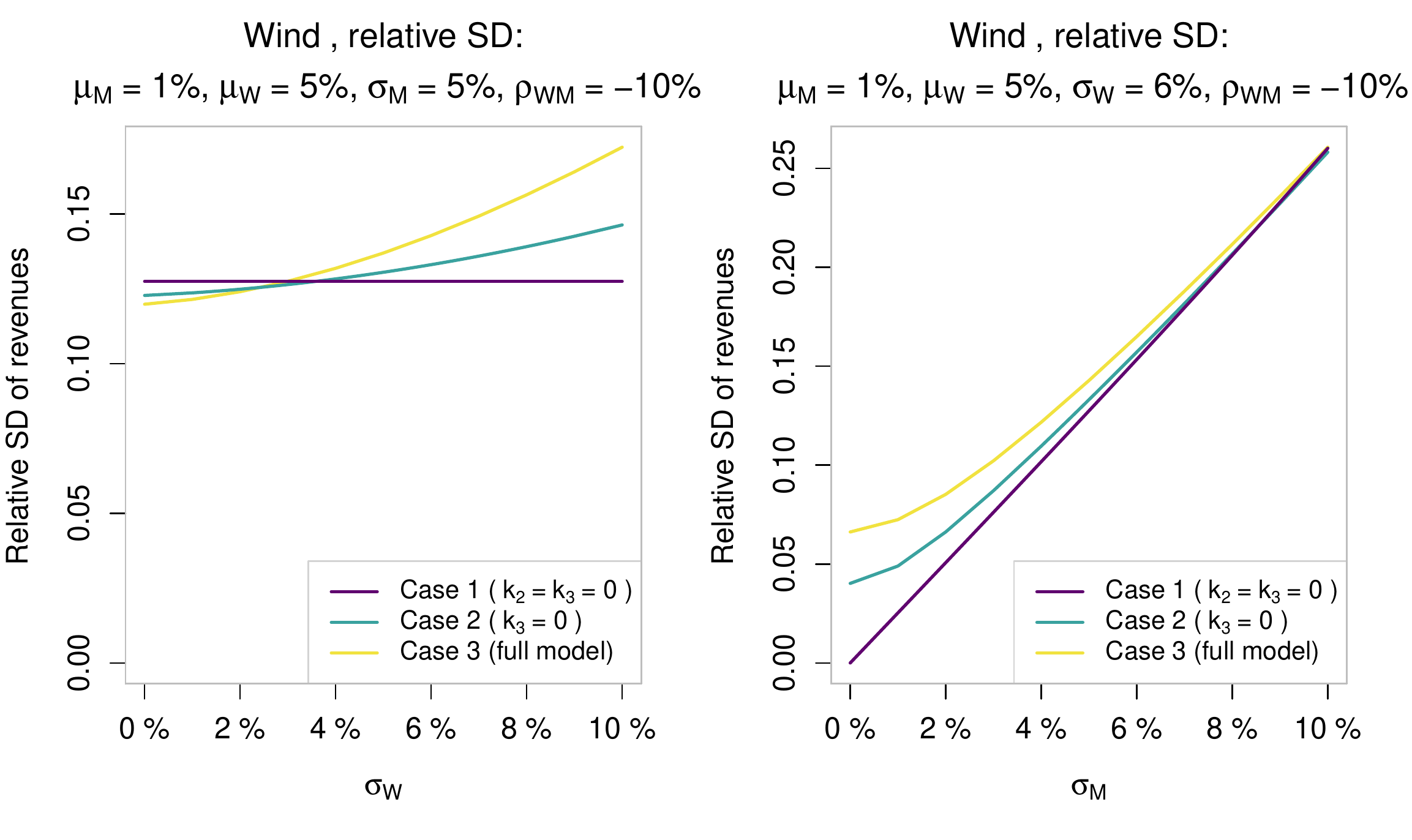}
\caption{Expected NPV (top, in euros) and standard deviation relative to the expected value (bottom) for 1~kW wind power investment. Expected values are presented for different values of $\mu_M$ and $\mu_W$ and standard deviations for $\sigma_M$ and $\sigma_W$ along the x-axes. The other parameter  values are given on top of each figure. Different colors indicate the full model (yellow), omitting cannibalization (teal), and omitting VRE merit-order effect and cannibalization (purple). The grey horizontal lines on the top panels indicate the NPV of investment and operating costs.}
\label{fig:NPVRisk}
\end{figure}

The expected revenues and risks for a wind power investment are presented in Figure \ref{fig:NPVRisk}. The results can be analyzed from two perspectives: 1) as the difference between \textbf{Cases 1}, \textbf{2} and \textbf{3} -- i.e. if considering or omitting the merit-order and cannibalization effects; and 2) the effect of subjective estimates about the future evolution of $\{W_t\}$ and $\{M_t\}$ on expected returns and risk in \textbf{Case 3}.

The top figures shows that the merit-order and cannibalization effects clearly affect the expected revenues, and these are especially amplified by a stronger VRE capacity growth rate (top left). This is intuitive, since a higher share of VRE in the generation mix leads to stronger cannibalization. The impact of these effects on risk is far less pronounced (the figures in the bottom), although this depends somewhat on the volatility parameters. Please see Figure \ref{fig:NPVRisk_variants} for further results regarding other parameters.

It is particularly noteworthy how the merit-order and cannibalization effects affect the profitability of the investment. In order to provide a point of comparison for the expected life-time revenue (per kW of wind power capacity), the top figures include an estimate of the NPV of investment- and O\&M costs (the grey line at $1800$ €/kW). We note that omitting the merit-order and cannibalization effects (\textbf{Case 1}) yields a profit which is independent of $\mu_W$ (top left). In addition, for \textbf{Case 1}, break-even is estimated to be reached even with a flattening MO curve (see $\mu_M<0$, top right). This means that any model that omits the cannibalization effect is prone to overestimate expected revenues. Such a model would deem an investment profitable under a relatively large range of subjective parameter values regarding the future, which -- in fact -- would render the investment unprofitable. 

When focusing on the full model, i.e. \textbf{Case 3}, it is evident that a strong future expansion of VRE capacity ($\mu_W$, top left) will eat the profits from a current investment, while the volatility ($\sigma_W$, bottom left) has a significant effect on the risks, especially in the case of high volatility of VRE capacity. Yet, the changes related to the slope of the MO curve ($\mu_M$ and $\sigma_M$, top and bottom right) have even more impact on expected revenues and risk within the assessed parameter range. This is intuitive, as the MO curve is the primary determinant for price formation. Any changes in the MO curve will directly translate into prices and revenues. 


Further illustration of expected revenues and risk with respect to all random walk parameters is provided in Figure \ref{fig:NPVRisk_variants}. This figure highlights that also the other parameters not presented in Figure \ref{fig:NPVRisk} do affect expected revenues and risk considerably, but omitting the merit-order and cannibalization effects does not make substantial difference with regard to some parameters (e.g. the effect of $\mu_M$ on risk).

\section{Investment timing}
\label{sec:investment_timing}
Since the VRE capacity level and the slope of the MO curve develop over time, investment may be postponed until such market conditions are sufficiently favorable. Thus, we address the problem of investment timing and how it is influenced by price impact and cannibalization effect. 

\subsection{Derivation}
The investment timing problem relies on the expected net present value derived in the previous sections. At time $t$ the expected net present value of revenues over the lifetime of the plant is:
\begin{equation}
V(W_t,M_t)=M_t(a-bW_t),\nonumber
\end{equation}
with constants 
\begin{align}
a=k_1\frac{1-e^{-(\beta-\mu_M) T}}{\beta-\mu_M}, \ b= (k_{2}+k_{3}) \frac{1-e^{-(\beta-\mu_{WM}) T}}{\beta-\mu_{WM}}.\nonumber
\end{align}
Thus, we express the expected value as a function of the VRE capacity level and slope of the MO curve which both develop stochastically over time. For known values of the capacity and the slope at time $t$, $W_t=w_t$ and $M_t=m_t$, $V(W_t,M_t)$ is known. However, at time $0$, $V(W_t,M_t)$ is random for $t>0$. 

We consider an infinite option to defer investment until the VRE penetration and an accompanying development of the MO curve justifies it. Clearly, this is a bi-variate real options problem. For known values of the capacity and the slope at time $0$, $W_0=w_0$ and $M_0=m_0$, the investment timing problem is 
\begin{equation}
F(w_0,m_0)=\max_{\tau}\{E_{W,M}^{w_0,m_0}[e^{-\beta \tau}(V(W_{\tau},M_{\tau})-I)\},\label{investtime}
\end{equation}
i.e., the problem is to determine the time $\tau$ at which the expected net present value is maximized (provided such $\tau$ exists). $I$ denotes the investment costs. The solution to the investment timing problem is derived in  \ref{app:RO}.

Consider the case of $k_2+k_3>0$ and thus $b>0$. We assume the existence of interdependent thresholds $(W^*,M^*)$ for the VRE capacity and the MO slope such that investment is optimal the first time $(W_t,M_t)=(W^*,M^*)$, where $W^*,M^*,\alpha_W,\alpha_M$ solve 
\begin{eqnarray}\label{eq:quadrMW}
\frac{1}{2}\Big(2\rho_{WM}\sigma_W\sigma_M\alpha_W\alpha_M+\sigma_W^2\alpha_W(\alpha_W-1)+\sigma_M^2\alpha_M(\alpha_M-1)\Big)+\mu_W\alpha_W+\mu_M\alpha_M-\beta=0
\end{eqnarray}
and
\begin{align}
W^*=-\frac{a}{b}\cdot\frac{\alpha_W}{\alpha_M-\alpha_W}, \  M^*=\frac{I}{a}\cdot\frac{\alpha_M-\alpha_W}{\alpha_M-1}.\label{eq:threshMW}
\end{align}
Thus, for fixed $W_t$ investment is optimal for $M_t\geq M^*(W_t)$, or for fixed $M_t$ investment is optimal for $W_t\leq W^*(M_t)$. Here, we express the threshold for the MO slope as a function of the VRE capacity. The expression of the threshold for VRE capacity as a function of MO slope can be found in \ref{app:RO}.

For an observed level of VRE capacity, i.e.\ fixed $W$, $\alpha_M$ solves
\begin{eqnarray}\label{eq:quadrM}
-\frac{1}{2}\Big(2\rho_{WM}\sigma_W\sigma_M\Big(\frac{Wb}{a-Wb}\Big)-\sigma_W^2\Big(\frac{Wb}{a-Wb}\Big)^2-\sigma_M^2\Big)\alpha_M(\alpha_M-1)&\\
+\Big(\frac{1}{2}\Big(-2\rho_{WM}\sigma_W\sigma_M\Big(\frac{Wb}{a-Wb}\Big)+\sigma_W^2\Big(\frac{Wb}{a-Wb}\Big)\Big(\frac{a}{a-Wb}\Big)\Big)-\mu_W\Big(\frac{Wb}{a-Wb}\Big)+\mu_M\Big)\alpha_M-\beta&=0\nonumber
\end{eqnarray}
and 
\begin{align}
M^*(W)=\frac{I}{a-Wb}\cdot\frac{\alpha_M}{\alpha_M-1}.\label{eq:threshM}
\end{align}





Consider next $k_2+k_3=0$ and thus $b=0$. The expected value of the plant no longer depends on the VRE capacity and the MO curve and the investment timing problem becomes univariate. Investment is optimal the first time $M_t=M^*$, where $\alpha_M$ solves 
\begin{eqnarray}
\frac{1}{2}\sigma_M^2\alpha_M(\alpha_M-1)+\mu_M\alpha_M-\beta=0\nonumber
\end{eqnarray}
and
\begin{align}
M^*=\frac{I}{a}\cdot\frac{\alpha_M}{\alpha_M-1}.\nonumber
\end{align}

\subsection{Numerical example}
\label{Numerical_example_investmenttiming}

\begin{figure}[!htbp]
\centering
\begin{minipage}{0.45\linewidth}
    \includegraphics[width=0.9\linewidth]{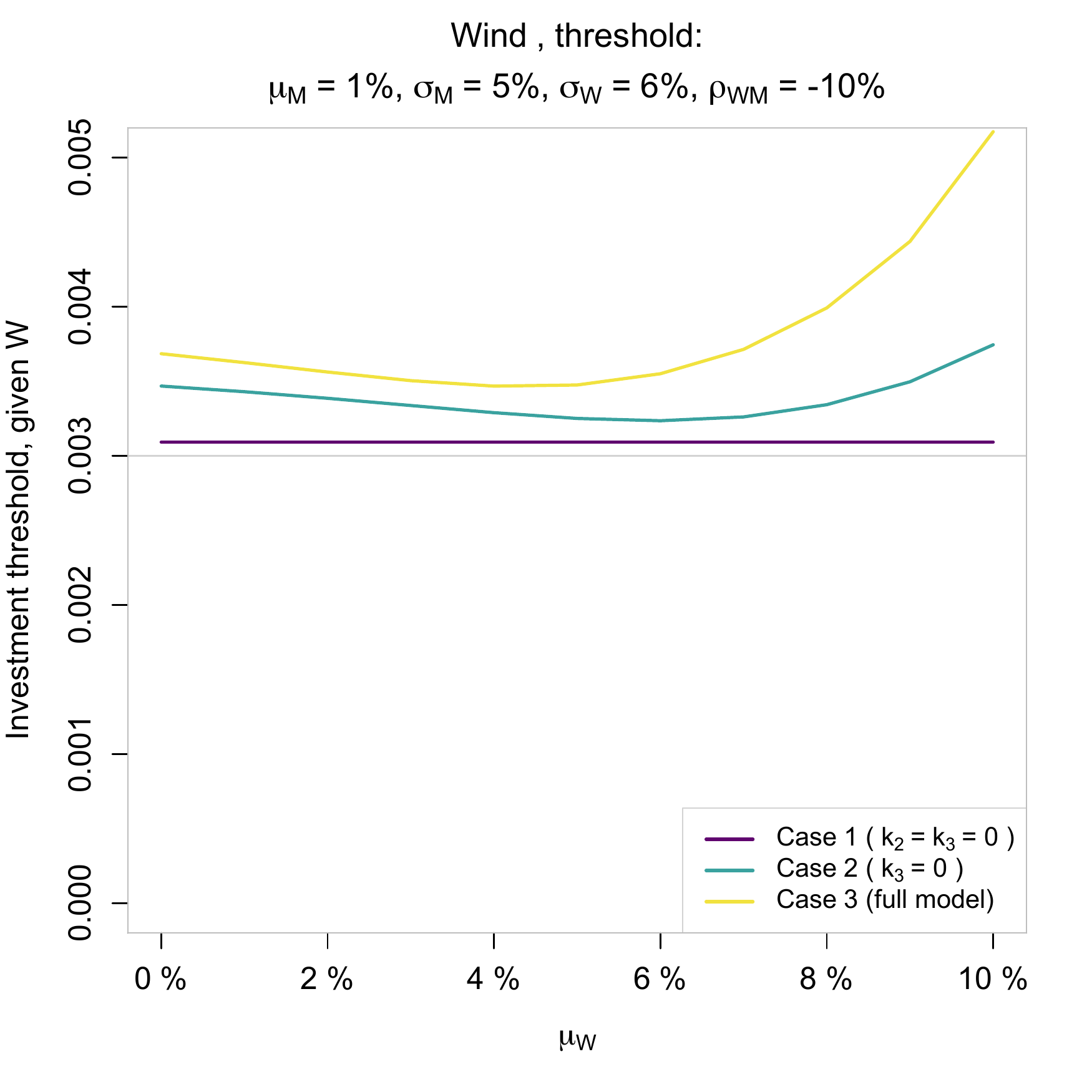}
\end{minipage}
\begin{minipage}{0.45\linewidth}
    \includegraphics[width=0.9\linewidth]{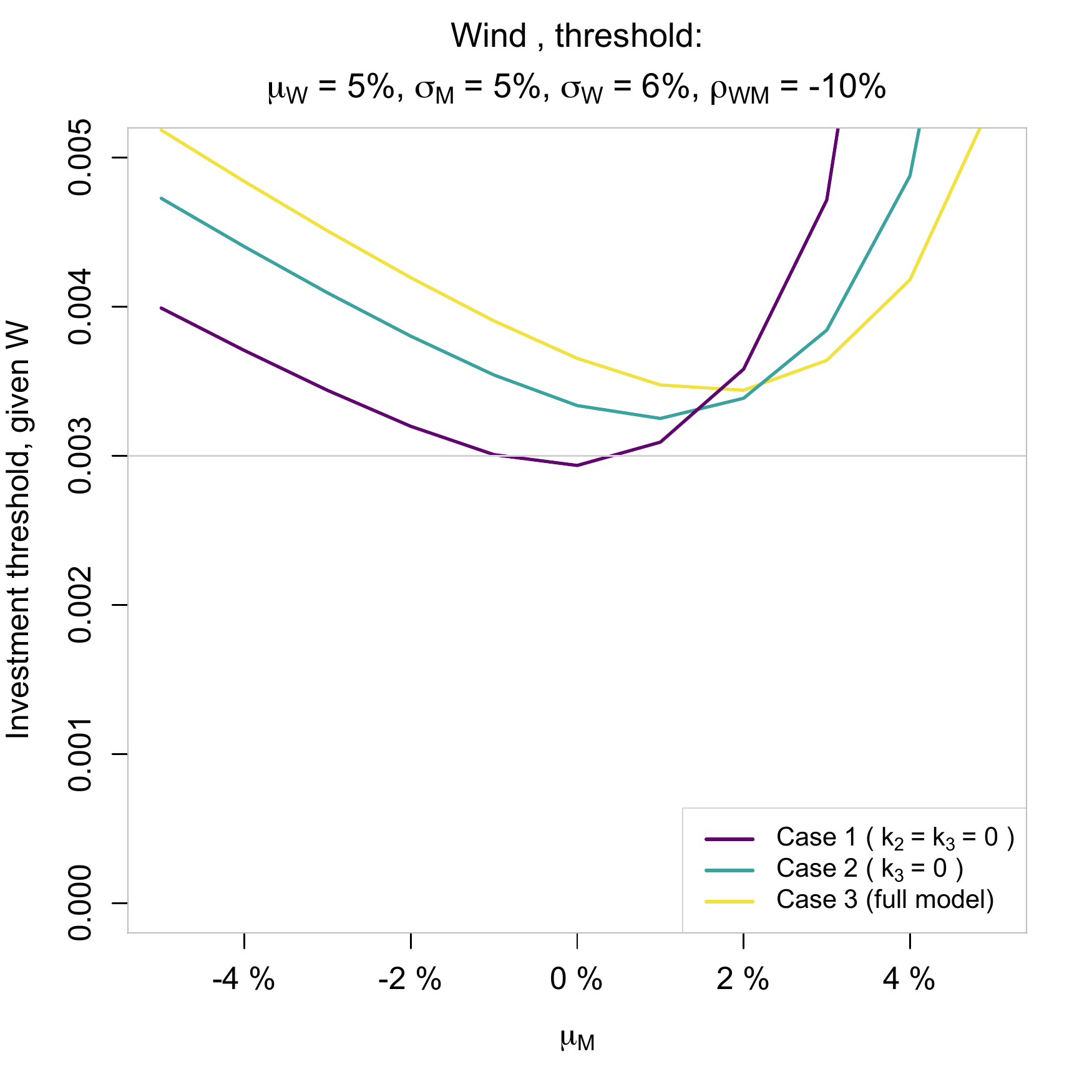}
\end{minipage}
\begin{minipage}{0.45\linewidth}
    \includegraphics[width=0.9\linewidth]{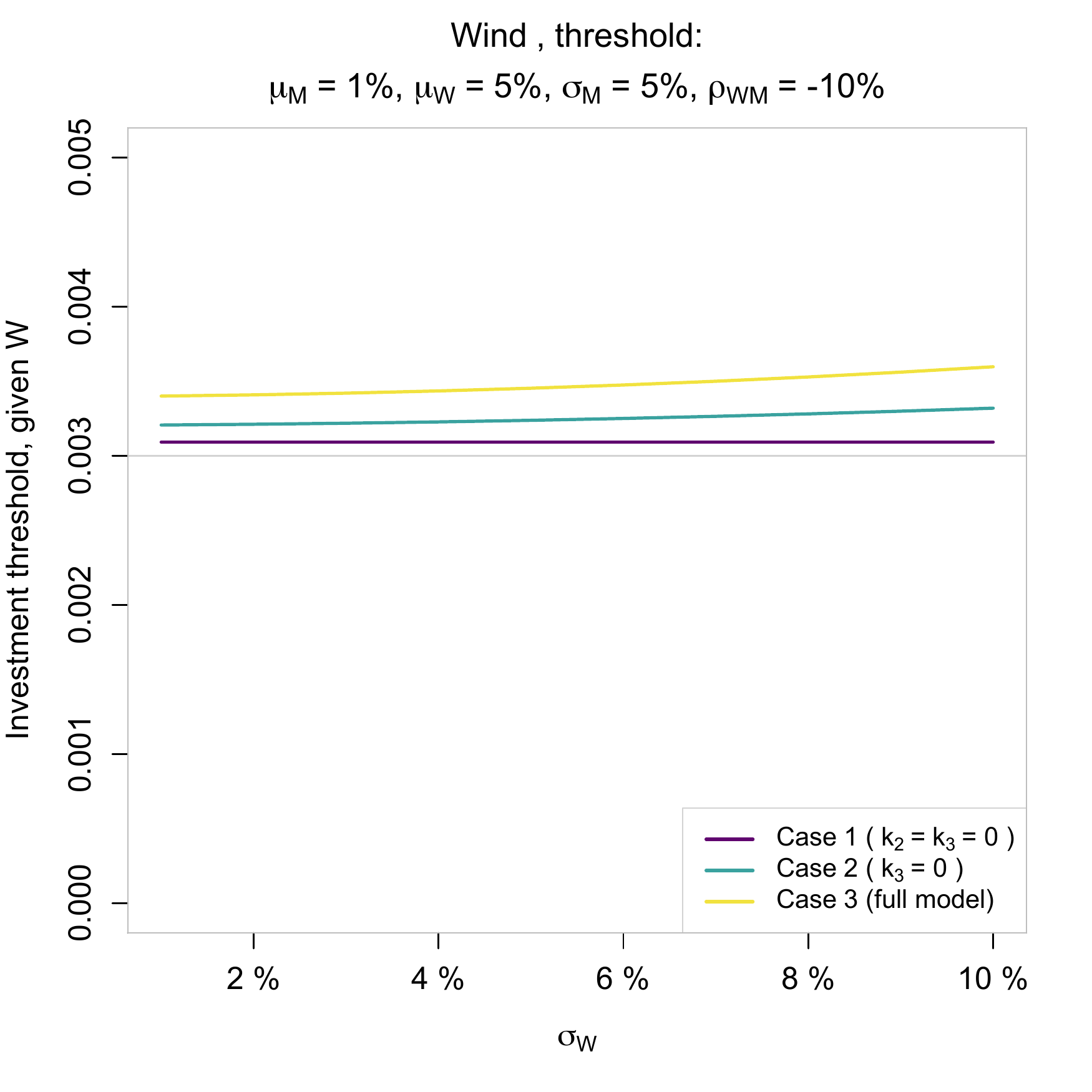}
\end{minipage}
\begin{minipage}{0.45\linewidth}
    \includegraphics[width=0.9\linewidth]{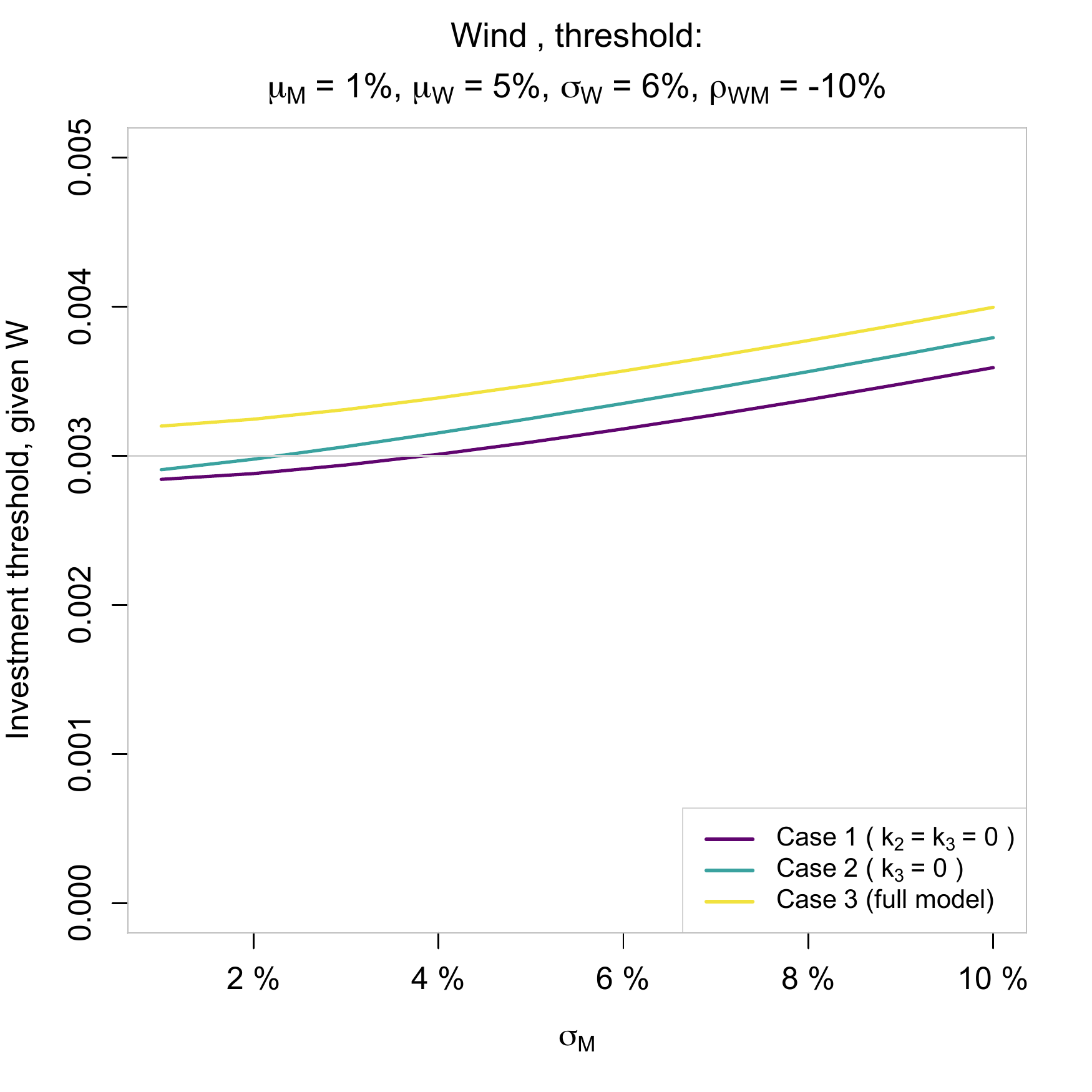}
\end{minipage}
\caption{Investment thresholds for the slope of the MO curve given VRE capacity level (in euros) for 1~MW wind power investment. Thresholds are presented for different values of $\mu_M$ and $\mu_W$ and standard deviations for $\sigma_M$ and $\sigma_W$ along the x-axes. Other parameters are given on top of each figure. Different colors indicate the full model (yellow), omitting cannibalization (teal), and omitting VRE merit-order effect and cannibalization (purple). The grey horizontal lines indicate the initial slope of the MO curve, $m_0$.}
\label{fig:Timing}
\end{figure}

Figure \ref{fig:Timing} shows the investment thresholds for the slope of the MO curve for a given level of VRE capacity and without merit order and cannibalization effects (\textbf{Case 1}), without cannibalization effects (\textbf{Case 2}) and with both effects (\textbf{Case 3}), respectively. Investment is optimal when the slope is at or above its threshold. For reference, the initial slope of the MO curve is also shown, revealing that for most parameter values immediate investment is not optimal but investment should be postponed. 

In the absence of merit order and cannibalization effects (top right, purple line), the impact of an increase in the drift of the MO slope can be divided into two. On the one hand, it will increase the value of the project, see also the previous section, requiring a lower MO slope at the time of investment to achieve the same project value. On the other hand, it will also raise the value of the option, and thus, require a higher MO slope to justify investment. The former effect dominates for low VRE growth rates and vice versa for high rates. The effect prevails in the presence of merit order and/or cannibalization effects (teal and yellow lines), although the threshold curves shifts upward and to the right. With these effects, the increase in the project as a result of an increase in drift is less pronounced, and hence, when the change in project value dominates the threshold is higher (teal and yellow lines are above purple line). The increase in the value of the option is likewise less pronounced, and when the change in option value dominates the threshold is lower (teal and yellow lines are below purple line). We conclude that by ignoring merit order and cannibalization effects, investment rates may for some drifts of the MO slope be too low, for others too high.

Similarly, the impact of an increase in the drift of the VRE capacity (top left) can be divided into two. It will decrease the project value and the option value, justifying a higher and lower capacity to trigger investment, respectively. As a result, the threshold is decreasing for some drifts and increasing for other. By ignoring any feedback of VRE on the price, however, the threshold is unaffected by the growth in VRE capacity. Without the merit order and/or cannibalization effects, the threshold for this numerical study will be too low and investment rates will be too high. 

As expected, an increase in volatility will raise the option value and thereby also the threshold, i.e.\ increasing uncertainty slows down investment. The threshold is increasing in the volatility of both the MO curve and the VRE capacity. In particular, the investor will postpone investment to wait for a steeper MO curve or a lower VRE capacity (although a positive drift of VRE capacity makes this less likely), and consequently higher electricity prices. By ignoring the merit order and/or cannibalization effects, investment rates will always be too high. As for the growth, by ignoring any feedback of VRE on the price, investment rates are unaffected by the volatility of VRE capacity.     

For comparison, Figure \ref{fig:Timing_npv} in \ref{sec:additional_results} shows the thresholds if the investment decision is based on the break-even according to net present value (NPV) rather than the real options value (ROV). According to the NPV rule, immediate investment is optimal if the NPV is positive. The NPV threshold is always less than the ROV threshold. In fact, it can be seen that NPV suggests immediate investment for many more parameter values than ROV, and so, there is a high risk of making poor investment decisions by disregarding the timing aspect. This is particularly pronounced in the presence of both the merit order and cannibalization effects. 

\section{Discussion and conclusion}
This paper introduces a framework to account for the merit order and cannibalization effects on VRE assets' revenues. These effects have been subject to extensive debate within the economic and engineering literature  \cite{HIRTH2013218, lamont2008assessing, winkler2016market, das2020learning, cannibal2018,mills2015strategies, brown2020decreasing}, but have been absent from assessments of investments in variable renewables. By conceptualizing the merit-order and cannibalization effects and providing analytic representations of them, we seek to incorporate observations from descriptive economics and energy system modelling into investment decisions.

While increasing uncertainty and deteriorating revenues over the life-time of real assets have already been addressed in an investment setting \cite{adkins2011renewing}, we specify these mechanisms in the case of VRE investment. Our stylized model accounts for the correlation between VRE penetration level and price, and yields closed-form expressions for the expected value and variance of revenues' net present value. Moreover, we illustrate how to use the analytical model in a real options framework for an investment timing problem. Previous literature on uncertainties for VRE investments \cite{5281848, kinias2017investment, kumbarouglu2008real, yang2008evaluating, RITZENHOFEN201676, boomsma2012renewable,boomsma2015market, kitzing2014risk, adkins2016subsidies} has not considered the cannibalization effect in the assessment of revenues. 


Our numerical examples quantify how the future evolution of market conditions, i.e.\ the capacity of VRE and the cost structure of dispatchable plants, could affect the expected revenue and risk of VRE investment. We show that the merit-order and cannibalization effects indeed have significant impacts on the revenues for VRE. Our estimated impact of cannibalization on short-term revenue ($-44$\% at a penetration level of around $30$\%) is corroborated by results of numeric simulations with more realistic models of the power market \cite{HIRTH2013218, brown2020decreasing}. 

For the impact on \emph{life-time} revenue and risk, we found no points of reference in the literature. Our examples show that cannibalization likewise has a significant impact on expected life-time revenue (Section \ref{Numerical_example_longterm}) and investment timing (Section \ref{Numerical_example_investmenttiming}). 
For instance, assuming that the future development of VRE capacity and MO curve of dispatchable plants in Poland are similar to the historical values for Germany ($\mu_W=10\%, \mu_M=1\%$), the project of our example (Figure \ref{fig:NPVRisk}) produces a loss of $800$ €/kW, i.e.\ $-40$\% return on investment. However, in the absence of MO and cannibalization effects, the investment calculation would yield a profit of $600$ €/kW, equalling a 33\% return on investment. Thus, disregarding the merit-order- and cannibalization effects can lead to greatly exaggerated revenue estimates, and therefore unprofitable investments. 

It should be noted, however, that cannibalization has a relatively smaller impact on life-time revenues ($-27$\%  at an end of lifetime penetration level of around $30$\% \footnote{As a comparison, the cannibalization effect for a 30\% penetration level of wind was estimated to around 50\%, both in our model and in the literature.}) than on instantaneous revenues, due to discounting. Thus, even though we observe substantial impact on lifetime revenues, the effect is less than the short-term estimates reported in the existing literature \cite{HIRTH2013218, lamont2008assessing, winkler2016market, das2020learning, cannibal2018,mills2015strategies, brown2020decreasing} might suggest. A higher discount rate than the modest 5\% applied here would further decrease the effect of cannibalization on estimated lifetime revenues.


 Another feature of the power market is the substantial uncertainty about the future development of the capacity mix, as well as the fuel and carbon prices, which in our model are represented through the random walks of VRE capacity and the MO curve. 
Our results indicate that the investor's belief about e.g.\ VRE capacity growth greatly affects the assessment of investment profitability and optimal timing. This may cause risk averse investors not to invest, thus causing a slower transition to a carbon neutral electricity system. Yet, a better understanding of the cannibalization effect can ultimately ensure a more adequate investment environment for VRE assets. From a policy maker's point of view, the understanding of such mechanisms can provide a rationale for compensating the losses and risks, e.g.\ by designing policy instruments, such as renewable energy auctions, which reduce the risk from market exposure. Also from a policy maker perspective, we further see that an increasing carbon tax, which gives rise to a steeper merit order curve of dispatchable plants, effectively mitigates the revenue decline due to cannibalization. A moderately increasing slope of the dispatchable merit order curve of 2\% per year, yields  a positive life-time profit even at a 5\% annual growth in VRE capacity.



There are some obvious limitations of our model, as stated already in Section \ref{sec:model_setup}. We disregard the seasonal variation of VRE generation, the variability of demand, and in particular the correlation between the demand and the VRE generation \footnote{The wind generation in Europe is for instance higher in winter, as is the demand.}. The model omits trade and electricity storage, the consideration of which would weaken the correlation between the investor's VRE generation and the net demand, and hence the price. The model is strictly valid only for VRE capacities less than or equal to the demand, as a higher VRE capacity can produce negative prices with the affine price function. In practical terms, this would allow VRE to generate only 20-40\% of annual energy. The impact of this limitation is further discussed in \ref{sec:validation}, showing a relatively minor effect in our calculated examples.

While our paper illustrates the potential usage of our closed-form expressions for an investment timing problem, we believe that the analytical framework may also be incorporated into other techniques. 
As we present closed-form expression for both expected revenues and variance (equations \eqref{eq:VRE_exp_revenue} and \eqref{eq:VRE_var_revenue}), the approach could be applied directly in modern portfolio theory. In this regard, further work is required to produce similar expressions for investments into dispatchable generation, allowing for a consistent portfolio model with a broader mix of generating technologies.

We conclude that:
\begin{itemize}
\item It is important to incorporate a representation of the cannibalization effect into capital budgeting methods for the electricity sector.
\item From the investor's perspective, doing so has a potentially large impact on both the expected return and the uncertainty regarding the expected return.
\item From the policy maker's perspective, awareness of the cannibalization effect may inform the choice of policy instruments towards those that (i) makes it more costly to emit CO$_2$ (such as a CO$_2$ tax) or (ii) reduce the risk relating to market exposure for VRE investors (such as green energy auctions).
\end{itemize}

\section*{Acknowledgments}

The work of Ekholm has been carried out with funding from the Academy of Finland (decision number 311010).

T.K.Boomsma acknowledges support from the project AHEAD, Analyses of Hourly Electricity Demand, funded by Energiteknologisk Udviklings- og Demonstrationsprogram (EUDP) under the Danish Energy Agency.

\appendix

\section{Additional data}
\label{Appdata}
To construct the linearization of the MO curve for dispatchable technologies in Section \ref{sec:model_setup}, we used data for capacities and compute the operational costs of the technologies present in the Polish electricity mix as of 2018. The EU-ETS price was assumed to have a starting value of $25$ €/tonne. The operational cost is the sum of fuel cost (Fuel cost / Conversion efficiency), the variable O\&M cost and the cost to emit CO$_2$ (EU-ETS price $\times$ emission factor). The resulting variable costs are listed in Table \ref{tbl:Operationalcosts_2018}.
\begin{table}[!hbp]
\caption{Capacities, conversion efficiencies and emission factors.}
\label{tbl:Capacities_2018}
\begin{center}
 \begin{tabular}{l c c c } 
 \hline
  Technology & Capacity (GW) &  Conversion efficiency & Emission factor (tonne CO$_2$/MWh) \\ 
 \hline
 Wind & 5.77 &  N/A & 0\\ 
 \hline
 Solar & 0.56 & N/A & 0\\
 \hline
 Hydro & 2.39  & N/A & 0\\
 \hline
 Lignite & 8.05 & 0.35 & 0.40\\
 \hline
 Hard coal & 19.20 & 0.40 & 0.34\\
 \hline
 Natural gas & 2.97 & 0.50 & 0.20\\
 \hline
\end{tabular}
\end{center}

\end{table}

\begin{table}[!hbp]
\caption{Fuel costs and variable Operation and maintenance (O\& M) costs.}
\label{tbl:Costs_2018}
\begin{center}
 \begin{tabular}{l c c} 
 \hline
  Technology & Fuel cost (€/MWh)  & Variable O\&M (€/MWh) \\ 
 \hline
 Wind & 0  & 0\\ 
 \hline
 Solar & 0 & 0\\
 \hline
 Hydro & 0 & 15 \\
 \hline
 Lignite & 3 & 5\\
 \hline
 Hard coal & 11.5 & 5\\
 \hline
 Natural gas & 25 & 6\\
 \hline
\end{tabular}
\end{center}

\end{table}

\begin{table}[hbt!]
\caption{Total operational costs for the technologies in the Polish electricity mix in 2018. The costs are computed using the fuel- and variable O\&M costs of Table \ref{tbl:Costs_2018} and conversion efficiencies and emissions factors from Table \ref{tbl:Capacities_2018}. The cost to emit CO$_2$ is set to $25$ €/tonne CO$_2$.}
\label{tbl:Operationalcosts_2018}
\begin{center}
 \begin{tabular}{l c} 
 \hline
  Technology & Total operational cost (€/MWh)   \\ 
 \hline
 Wind & 0  \\ 
 \hline
 Solar & 0 \\
 \hline
 Hydro &  15 \\
 \hline
 Lignite & 42\\
 \hline
 Hard coal & 55\\
 \hline
 Natural gas & 66\\
 \hline
\end{tabular}
\end{center}

\end{table}

The values to compute the LCOE of wind [€/MWh] and the NPV of wind investment [€/kW] that are used for comparison with the revenues are computed with the values for investment cost, lifetime and discount rate in Table \ref{tbl:LCOE_data}.
\begin{table}[hbt!]
\caption{Data on costs and other parameters used to calculate LCOE and costs.}
\label{tbl:LCOE_data}
\begin{center}
 \begin{tabular}{l c c c c c} 
 \hline
  Technology & Investment cost [€/kW]  &  Fixed O\&M [€/kW/year] & Lifetime [years] & Discount rate [\%] & Capacity factor \\ 
 \hline
 Wind & 1200  & 40 & 25 & 5 & 0.31\\
 \hline
 Solar & 800 & 30 & 25 & 5 & 0.15\\
 \hline
 
\end{tabular}
\end{center}
\end{table}

\section{Derivation of the variance of revenues}
\label{sec:AppVariance}

The variance of the net present value of revenues is given by
\begin{align}
    \label{eq:VRE_var_revenue_start}
    Var(w_0,m_0) &= E_{G,W,M}^{w_0,m_0}\left[ \left( \int_0^T e^{-\beta t} R(G_t,W_t,M_t) dt \right)^2 \right] - E_{G,W,M}^{w_0,m_0}\left[\int_0^T e^{-\beta t} R(G_t,W_t,M_t) dt\right]^2
\end{align}

The last term is the square of \eqref{eq:VRE_exp_revenue}, so it remains to find an expression for the first term:
\begin{align}
    \label{eq:VRE_var_revenue_2ndterm}
    & E_{G,W,M}^{w_0,m_0}\left[ \left( \int_0^T e^{-\beta t} R(G_t,W_t,M_t) dt \right)^2 \right] \\
    &= E_{G,W,M}^{w_0,m_0}\left[ \int_0^T e^{-\beta t} R(G_t,W_t,M_t) dt \int_0^T e^{-\beta s} R(G_s,W_s,M_s) ds \right] \nonumber \\
    &= E_{G,W,M}^{w_0,m_0}\left[ \int_0^T \int_0^T e^{-\beta (t+s)} R(G_t,W_t,M_t) R(G_s,W_s,M_s) ds dt  \right] \nonumber \\
    &= 2 \int_0^T \int_t^T e^{-\beta (t+s)} E_{G,W,M}^{w_0,m_0}\left[R(G_t,W_t,M_t) R(G_s,W_s,M_s)\right] ds dt   \nonumber \\
    &= 2 \int_0^T \int_t^T e^{-\beta (t+s)} E_{W,M}^{w_0,m_0}\left[ (k_1 M_t-(k_2+k_3) M_t W_t)(k_1 M_s-(k_2+k_3) M_s W_s) \right] ds dt   \nonumber \\
    &= 2 \int_0^T \int_t^T e^{-\beta (t+s)} E_{W,M}^{w_0,m_0}\left[ \left( k_1^2 M_t M_s 
    - k_1 (k_2+k_3) \left( M_tM_s W_s + M_t M_s W_t \right)
    + (k_2+k_3)^2 M_t M_s W_t W_s \right) \right] ds dt  \nonumber \\
    &= 2 \int_0^T \int_t^T e^{-\beta (t+s)}  \Big( 
    \begin{aligned}[t] 
    & k_1^2E_{M}^{m_0}\left[M_t M_s\right]  
    - k_1 (k_2+k_3) \left( E_{W,M}^{w_0,m_0}\left[M_t W_sM_s\right] + E_{W,M}^{w_0,m_0}\left[W_tM_t M_s\right] \right) \\
    & + (k_2+k_3)^2 E_{W,M}^{w_0,m_0}\left[W_tM_t W_sM_s\right] \Big) ds dt 
    \end{aligned} \nonumber
\end{align}

The fourth line takes advantage of integrands' symmetry between $t$ and $s$. The fifth line takes the expectation with regard to $G_t$, using the fact that $G_t$ and $G_s$ are independent almost everywhere, i.e. when $t \neq s$. 

Next, we need expressions for the expected values on the last line of \eqref{eq:VRE_var_revenue_2ndterm}.
Let $N_t=\ln(M_t)$, i.e. a  Brownian motion with drift.
Using this, $E_{M}^{m_0}\left[M_t M_s\right] = E_{M}^{m_0}\left[e^{N_t + N_s}\right]$, which equals the first moment generating function for the random variable $N_t + N_s$.
This can be split as  $N_t + N_s = 2 N_t + (N_s - N_t)$. 
As $s \geq t$, the two terms are independent, and 
$E_{M}^{m_0}\left[e^{N_t + N_s}\right] = E_{M}^{m_0}\left[e^{2 N_t}\right] E_{M}^{m_0}\left[e^{N_s - N_t}\right]$.
This gives
\begin{align}
E_{M}^{m_0}\left[M_t M_s\right] = m_0^2 \, e^{(2\mu_M + \sigma_M^2)t} \, e^{\mu_M (s-t)} = m_0^2 \, e^{(\mu_M + \sigma_M^2) t + \mu_M s}
\end{align}

Similarly, let  $U_t=\ln(W_t)$. Then $E_{W,M}^{w_0,m_0}\left[W_t M_t M_s\right] = E_{W,M}^{w_0 m_0}\left[e^{U_t+N_t+N_s}\right] = E_{W,M}^{w_0 m_0}\left[e^{U_t+2N_t}\right]E_{M}^{m_0}\left[e^{N_s-N_t}\right]$,
based on the independence of $U_t+2N_t$ and $N_s-N_t$. Using this, and taking into account the dependence between $U_t$ and $2N_t$,
\begin{align}
E_{W,M}^{w_0,m_0}\left[W_t M_t M_s\right] = m_0^2 w_0 \, e^{(2\mu_M + \mu_W + 2 \rho_{WM} \sigma_W \sigma_M + \sigma_M^2)t} \, e^{\mu_M (s-t)} = m_0^2 w_0 \, e^{(\mu_{WM} + \rho_{WM} \sigma_W \sigma_M + \sigma_M^2) t + \mu_M s}
\end{align}

Similar results hold for $E_{W,M}^{w_0,m_0}\left[M_t W_s M_s\right]$ and $E_{W}^{w_0}\left[W_tM_t W_sM_s\right]$, yielding the following expressions that were required for \eqref{eq:VRE_var_revenue_2ndterm}:
\begin{align}
\label{eq:VRE_var_expectations}
    E_{W,M}^{w_0,m_0}\left[M_t M_s \right] &= m_0^2 \, e^{(\mu_M+\sigma_M^2) t + \mu_M s}  \\
    E_{W,M}^{w_0,m_0}\left[M_t W_s M_s \right] &= m_0^2 w_0 \, e^{(\mu_{WM} + \rho_{WM} \sigma_W \sigma_M + \sigma_M^2) t + \mu_M s} \nonumber \\
    E_{W,M}^{w_0,m_0}\left[W_t M_t M_s \right] &= m_0^2 w_0 \, e^{(\mu_M + \rho_{WM} \sigma_W \sigma_M + \sigma_M^2) t + \mu_{WM} s} \nonumber \\
    E_{W,M}^{w_0,m_0}\left[W_t M_t W_s M_s \right] &= m_0^2 w_0^2 \, e^{(\mu_{WM} + \sigma_{WM}^2) t + \mu_{WM} s} \nonumber
\end{align}

Let us define shorthands for the constants in the exponents, also accounting for the discount rate $\beta$:
\begin{align}
    \label{eq:VRE_var_shorthands}
    &  p_1 = \mu_M - \beta                                 \\
    &  p_2 = \mu_M - \sigma_M^2 - \beta                        \nonumber \\
    &  p_3 = \mu_{WM} - \beta                           \nonumber \\
    &  p_4 = \mu_{WM} + \rho_{WM} \sigma_W \sigma_M + \sigma_M^2 - \beta                   \nonumber \\
    &  p_5 = \mu_M + \rho_{WM} \sigma_W \sigma_M + \sigma_M^2 - \beta     \nonumber \\
    &  p_6 = \mu_{WM} + \sigma_{WM}^2 - \beta     \nonumber 
\end{align}

Using \eqref{eq:VRE_var_expectations} and \eqref{eq:VRE_var_shorthands} in \eqref{eq:VRE_var_revenue_2ndterm}, then inserting \eqref{eq:VRE_exp_revenue} and \eqref{eq:VRE_var_revenue_2ndterm} into \eqref{eq:VRE_var_revenue_start}, and finally solving the integrals gives a closed-form expression for the variance:
\begin{align}
    \label{eq:VRE_var_revenue_App}
    Var(w_0,m_0) = & 2 m_0^2 \Bigg( 
    \begin{aligned}[t] 
                 k_1^2 & \,\, \frac{p_2 + p_1 e^{(p_1+p_2) T} - (p_1+p_2) e^{p_1 T}}{p_1 p_2(p_1 + p_2)} \\ 
    - w_0k_1 (k_2+k_3) & \,\, \frac{p_4 + p_1 e^{(p_1+p_4) T} - (p_1+p_4) e^{p_1 T}}{p_1 p_4(p_1 + p_4)} \\ 
    - w_0k_1 (k_2+k_3) & \,\, \frac{p_5 + p_3 e^{(p_3+p_5) T} - (p_3+p_5) e^{p_3 T}}{p_3 p_5(p_3 + p_5)} \\ 
    + w_0^2(k_2+k_3)^2 & \,\, \frac{p_6 + p_3 e^{(p_3+p_6) T} - (p_3+p_6) e^{p_3 T}}{p_3 p_6(p_3 + p_6)} 
    \Bigg) 
    \end{aligned}
    \\ \nonumber &
    - m_0^2 \left( k_1\frac{1-e^{-p_1T}}{p_1} - w_0(k_2+k_3)\frac{1-e^{-p_3 T}}{p_3} \right)^2 
\end{align}

\section{Real options derivations}\label{app:RO}

With an infinite option, we obtain a time-homogeneous value process. As a result, the dynamic programming recursion of \eqref{investtime} is
\begin{eqnarray}
F(W,M)=\max\Big\{V(W,M),\frac{1}{1+\beta dt}E[F(W+dW,M+dM\mid W,M)\Big\}.\nonumber
\end{eqnarray}
where $W$ and $M$ refer to the current levels of VRE capacity and the MO slope, respectively. According to this recursion, at any point in time, the investor may undertake investment and realize its value or decide to defer. The decision depends on the trade-off between the current expected value of the plant and the discounted expected future value of the option to invest.

When it is optimal to defer investment, this means
\begin{eqnarray}
E[dF(W+dW,M+dM\mid W,M)=\beta F(W,M)dt.\nonumber
\end{eqnarray}
Using Ito's Lemma to expand the left-hand-side, we obtain the partial differential equation (PDE)
\begin{eqnarray}
\frac{1}{2}\Big(2\rho_{WM}\sigma_W\sigma_MWM\frac{\partial^2F}{\partial W\partial M}+\sigma_W^2W^2\frac{\partial^2F}{\partial W^2}+\sigma_M^2M^2\frac{\partial^2F}{\partial M^2}\Big)+\mu_WW\frac{\partial F}{\partial W}+\mu_MM\frac{\partial F}{\partial M}-\beta F=0.\nonumber
\end{eqnarray}
This is subject the boundary conditions 
\begin{align}
&F(W,M)=V(W,M)-I, \ \frac{\partial F}{\partial W}=-bM \ \frac{\partial F}{\partial M}=a-bW\nonumber\\[2mm]
&\lim_{M\rightarrow 0}F(W,M)=0, \ \lim_{W\rightarrow\infty}F(W,M)=0\nonumber
\end{align}
i.e., when it is optimal to invest, the net present value of the plant is realized, and when the slope of the merit order curve tends to zero or the VRES capacity to infinity, the investment is worthless. 

To obtain a solution to the PDE, we assume that $F(W,M)=AW^{\alpha_W}M^{\alpha_M}$ with $\alpha_W<0<1<\alpha_M$. Then, $\alpha_W, \alpha_M$ satisfy
\begin{align}
\frac{1}{2}\Big(2\rho_{WM}\sigma_W\sigma_M\alpha_W\alpha_M+\sigma_W^2\alpha_W(\alpha_W-1)+\sigma_M^2\alpha_M(\alpha_M-1)\Big)+\mu_W\alpha_W+\mu_M\alpha_M-\beta=0.\label{eq:quadr}
\end{align}
Moreover, at the boundary, $W,M$ and $A$ satisfy
\begin{align}
AW^{\alpha_W}M^{\alpha_M}=M(a-bW)-I,\label{eq:boundary}\\ \alpha_WAW^{\alpha_W-1}M^{\alpha_M}=-bM, \  \alpha_MAW^{\alpha_W}M^{\alpha_M-1}=a-bW.\nonumber
\end{align}

The solution to \eqref{eq:quadr} and \eqref{eq:boundary} is given in \eqref{eq:quadrMW} and \eqref{eq:threshMW}. Equivalently, for an observed level of VRE capacity , i.e.\ for fixed $W$, we refer to \eqref{eq:quadrM} and \eqref{eq:threshM}.

Alternatively, for an observed slope of the MO curve, i.e.\ for fixed $M$, $\alpha_W$ solves
\begin{eqnarray}
-\frac{1}{2}\Big(2\rho_{WM}\sigma_W\sigma_M\Big(\frac{I}{Ma-I}\Big)-\sigma_W^2-\sigma_M^2\Big(\frac{I}{Ma-I}\Big)^2\Big)\alpha_W(\alpha_W-1)&\nonumber\\
+\Big(\frac{1}{2}\Big(2\rho_{WM}\sigma_W\sigma_M-\sigma_M^2\Big(\frac{I}{Ma-I}\Big)\Big(\frac{Ma}{Ma-I}\Big)\Big)+\mu_W-\mu_M\Big(\frac{I}{Ma-I}\Big)\Big)\alpha_W&\nonumber\\
+\frac{1}{2}\sigma_M^2\Big(\frac{I}{Ma-I}\Big)\Big(\frac{Ma}{Ma-I}\Big)+\mu_M\Big(\frac{Ma}{Ma-I}\Big)-\beta&=0\nonumber
\end{eqnarray}
and 
\begin{align}
W^*(M)=-\frac{Ma-I}{Mb}\cdot\frac{\alpha_W}{1-\alpha_W}.\nonumber
\end{align}

\section{Additional numerical results}
\label{sec:additional_results}

\begin{figure}[!htb]
\centering
    \newpage
    \includegraphics[width=0.8\linewidth,clip=true]{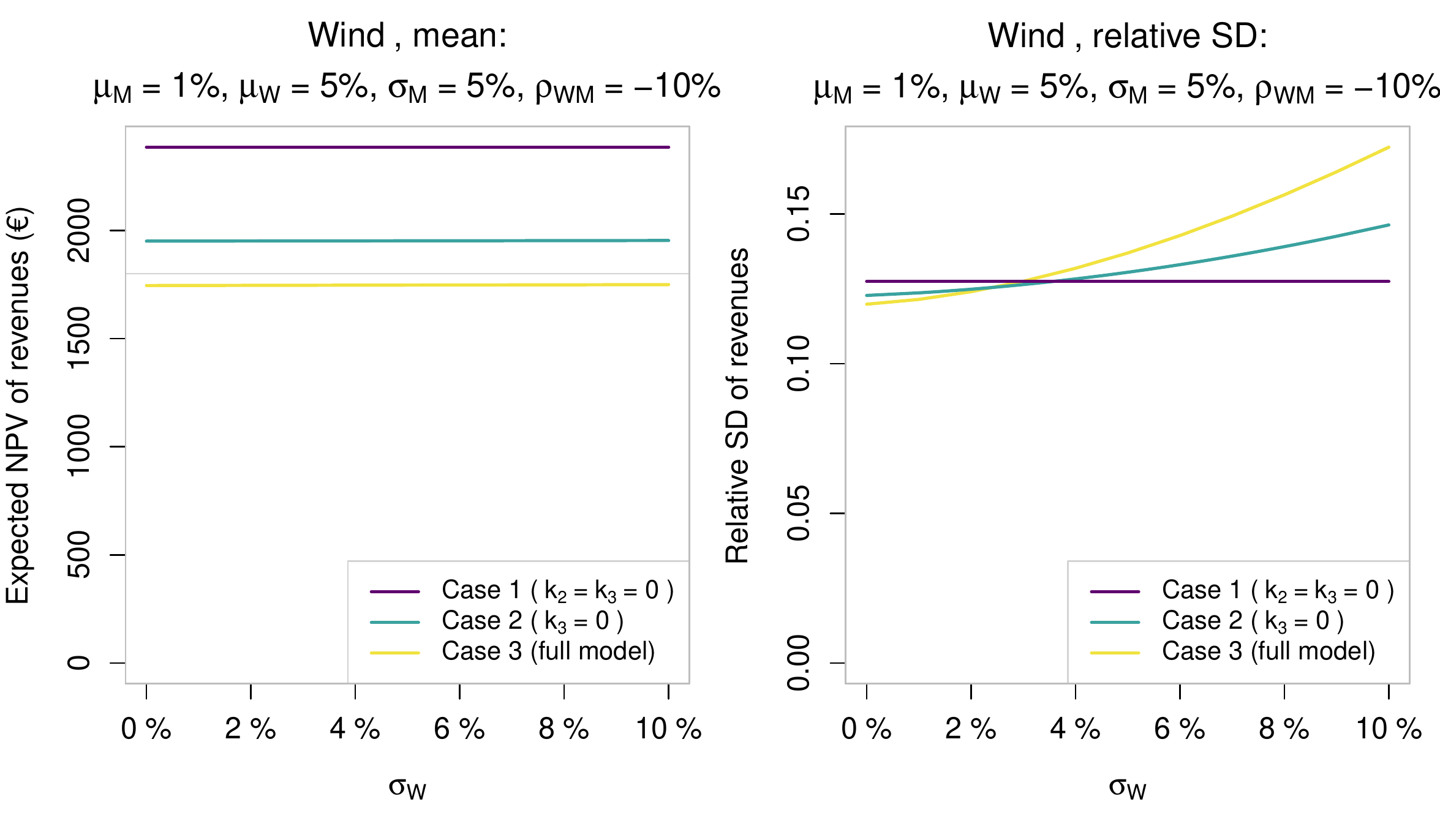}
    \includegraphics[width=0.8\linewidth,clip=true]{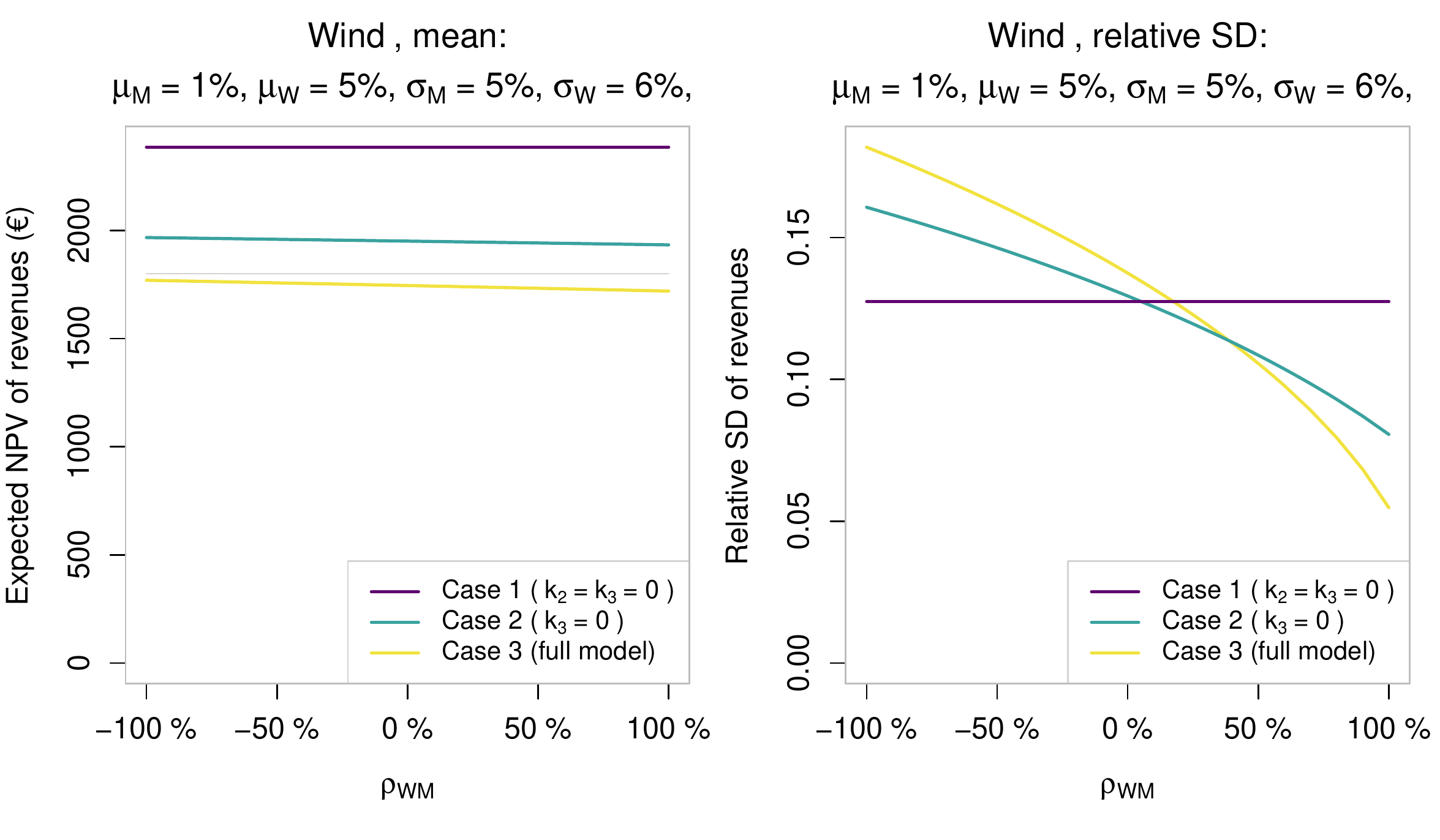}
\caption{The expected value (left) and standard deviation (right) of revenues' NPV for a 1 kW wind power investment with different values of $\mu_M$, $\mu_W$, $\sigma_M$, $\sigma_W$, $\rho_{MW}$ (x-axis).
Different colors indicate full model (yellow), omitting cannibalization (teal) and omitting VRE merit order effect and cannibalization (purple). 
 \label{fig:NPVRisk_variants}
}
\end{figure}

Figure \ref{fig:NPVRisk_variants} presents the expected revenues and risk as a function all the random walk parameters, thus expanding the results presented in Figure \ref{fig:NPVRisk}.

\begin{figure}[!htb]
\centering
    \includegraphics[width=0.8\linewidth,clip=true]{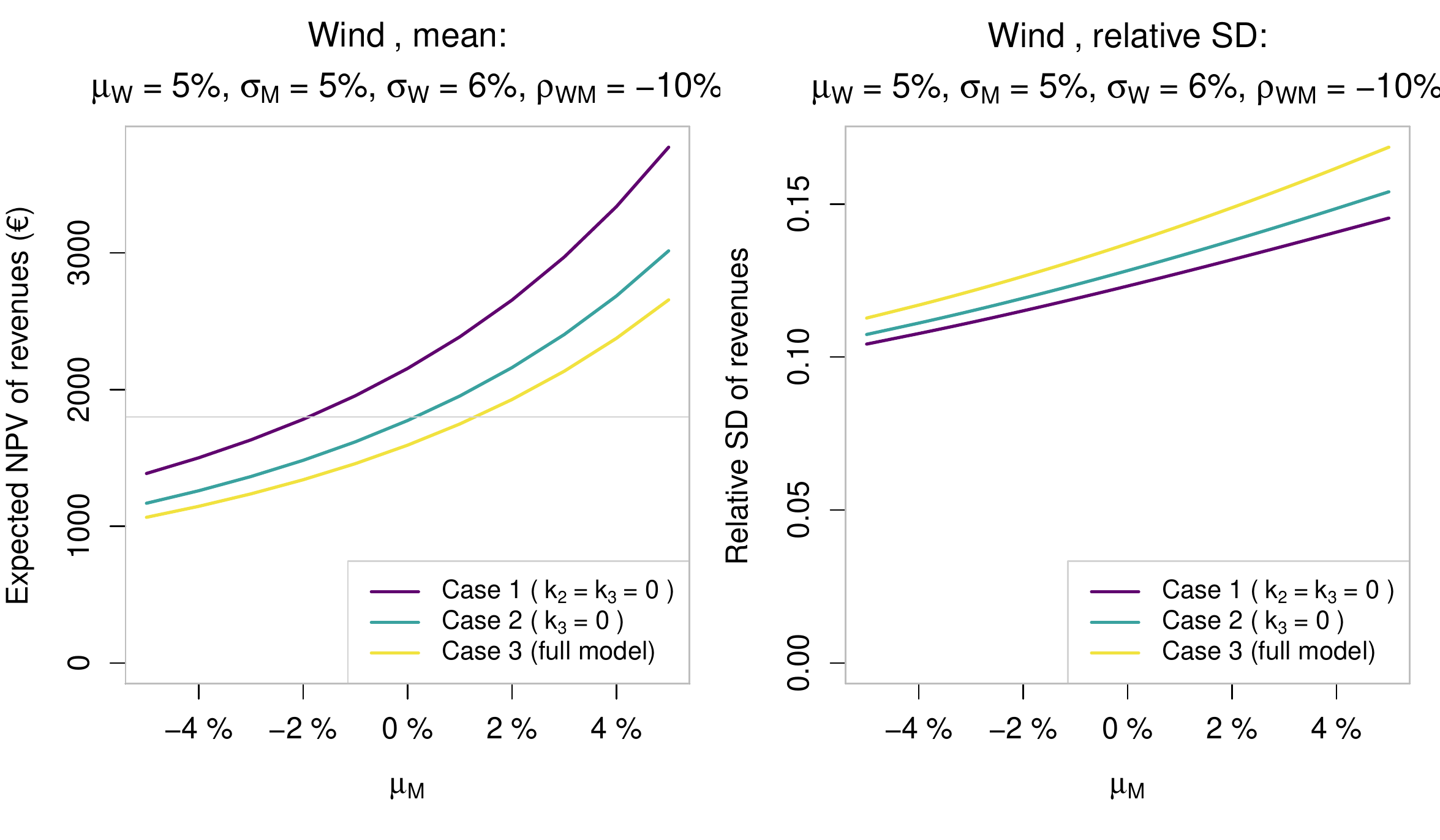}
    \includegraphics[width=0.8\linewidth,clip=true]{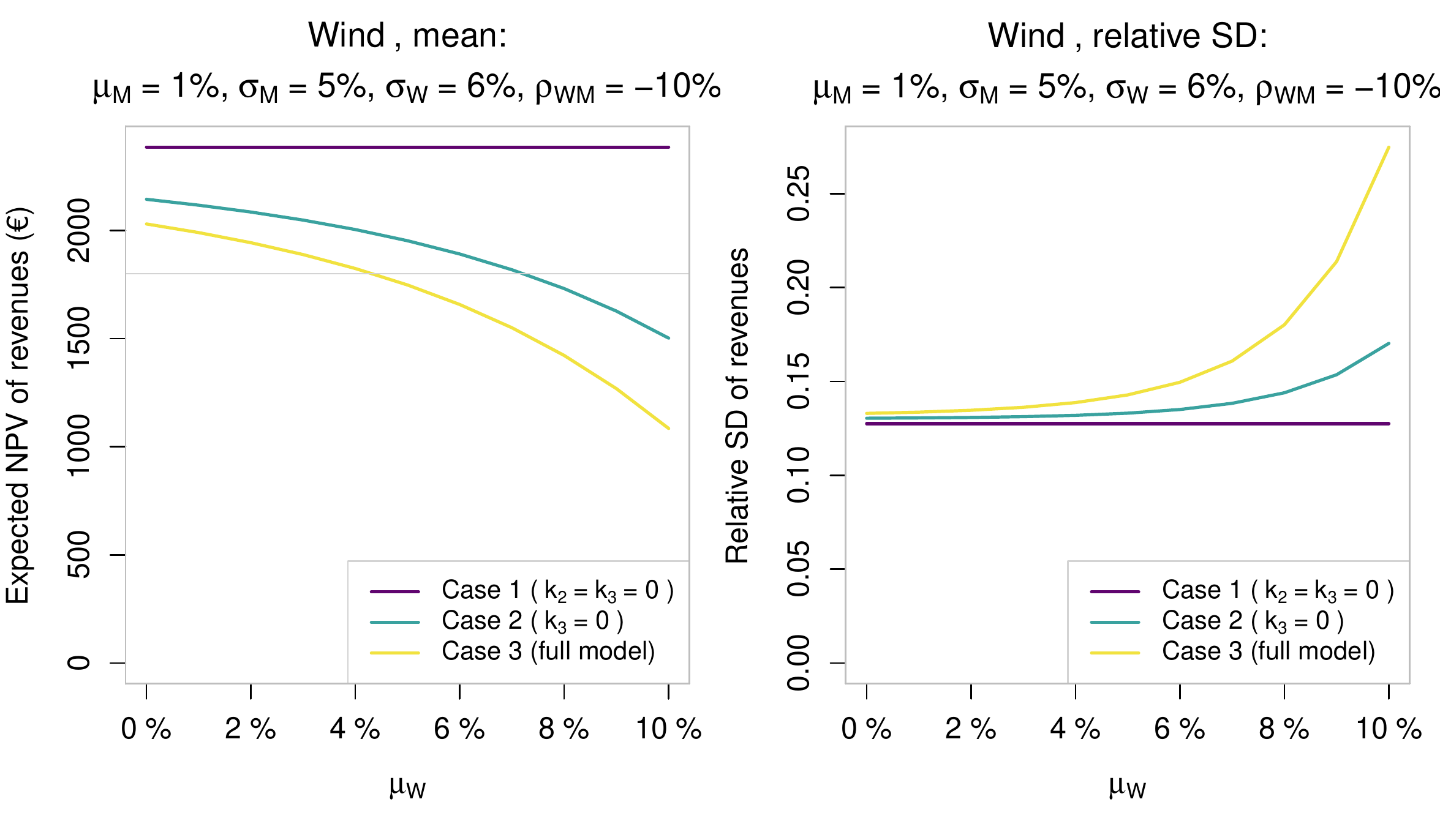}
    \includegraphics[width=0.8\linewidth,clip=true]{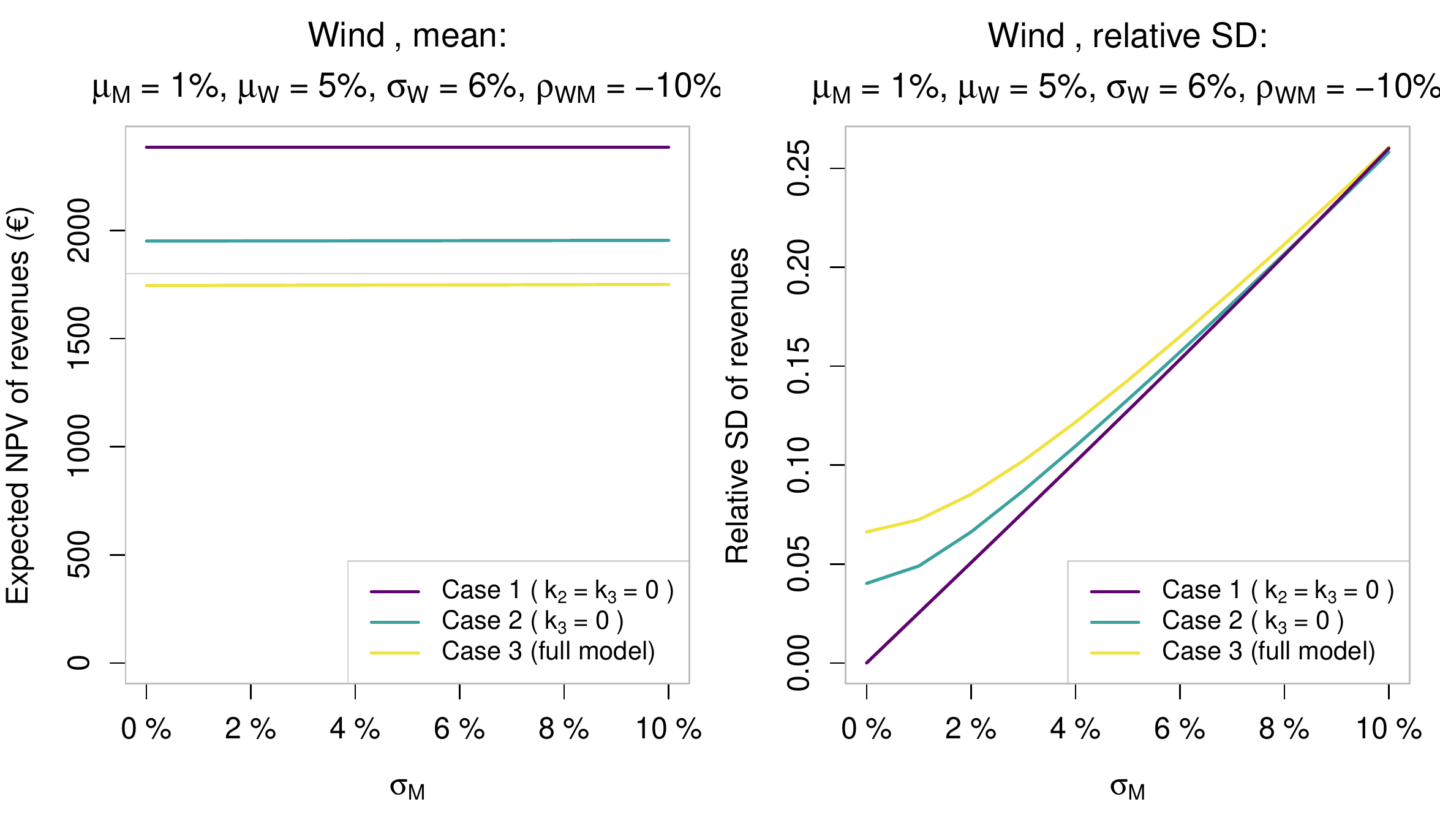}
\end{figure}

\begin{figure}[!htbp]
\centering
\begin{minipage}{0.45\linewidth}
    \includegraphics[width=0.9\linewidth]{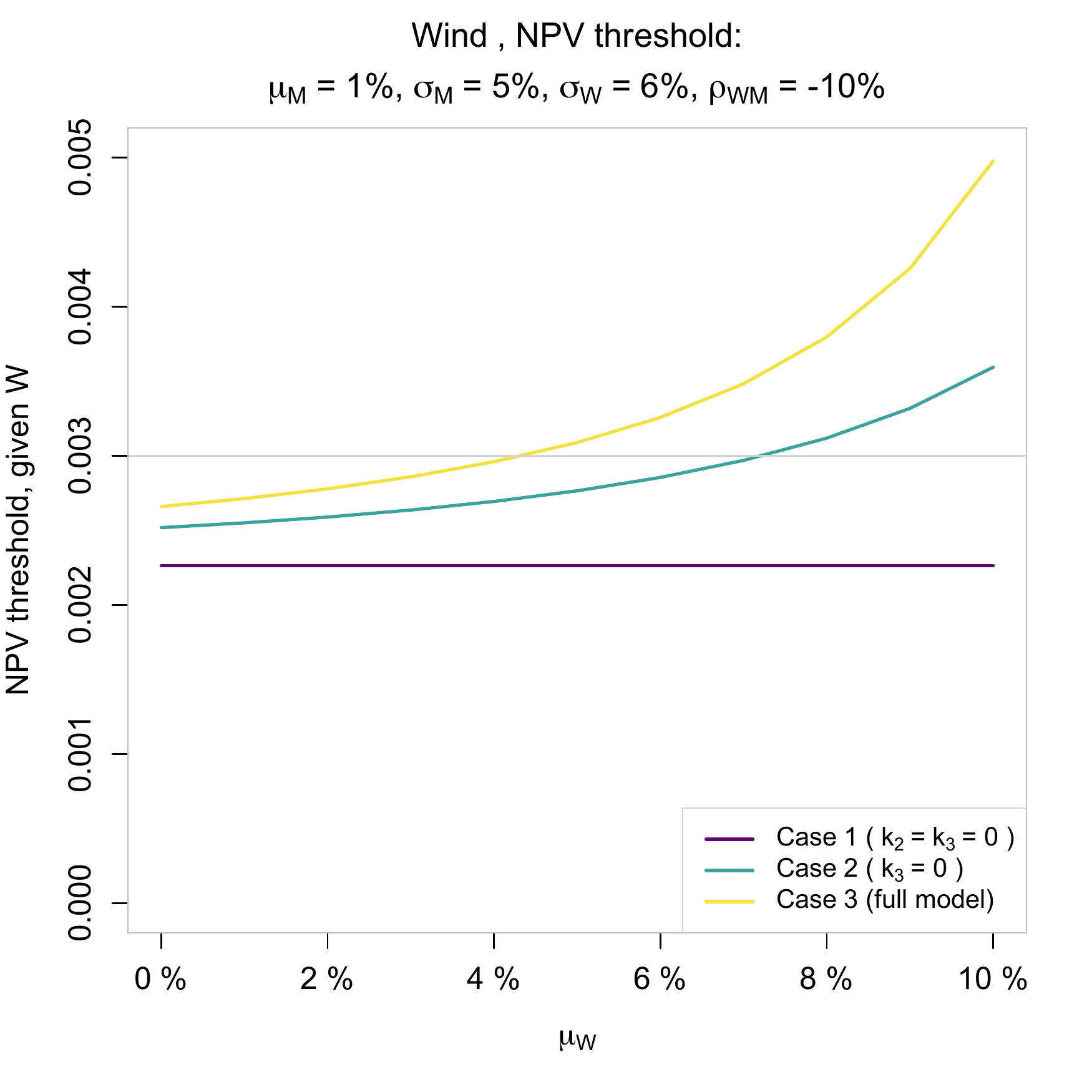}
\end{minipage}
\begin{minipage}{0.45\linewidth}
    \includegraphics[width=0.9\linewidth]{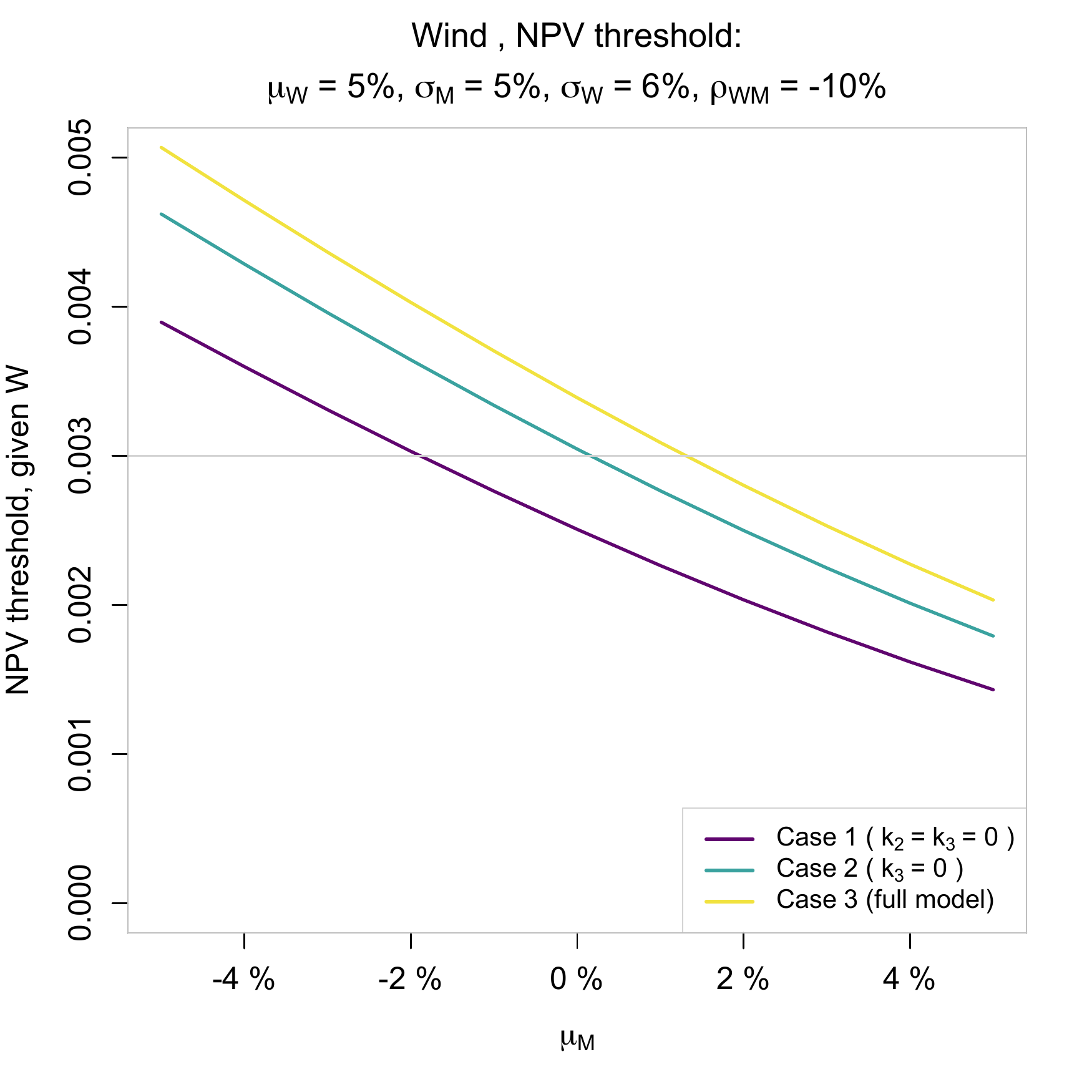}
\end{minipage}
\begin{minipage}{0.45\linewidth}
    \includegraphics[width=0.9\linewidth]{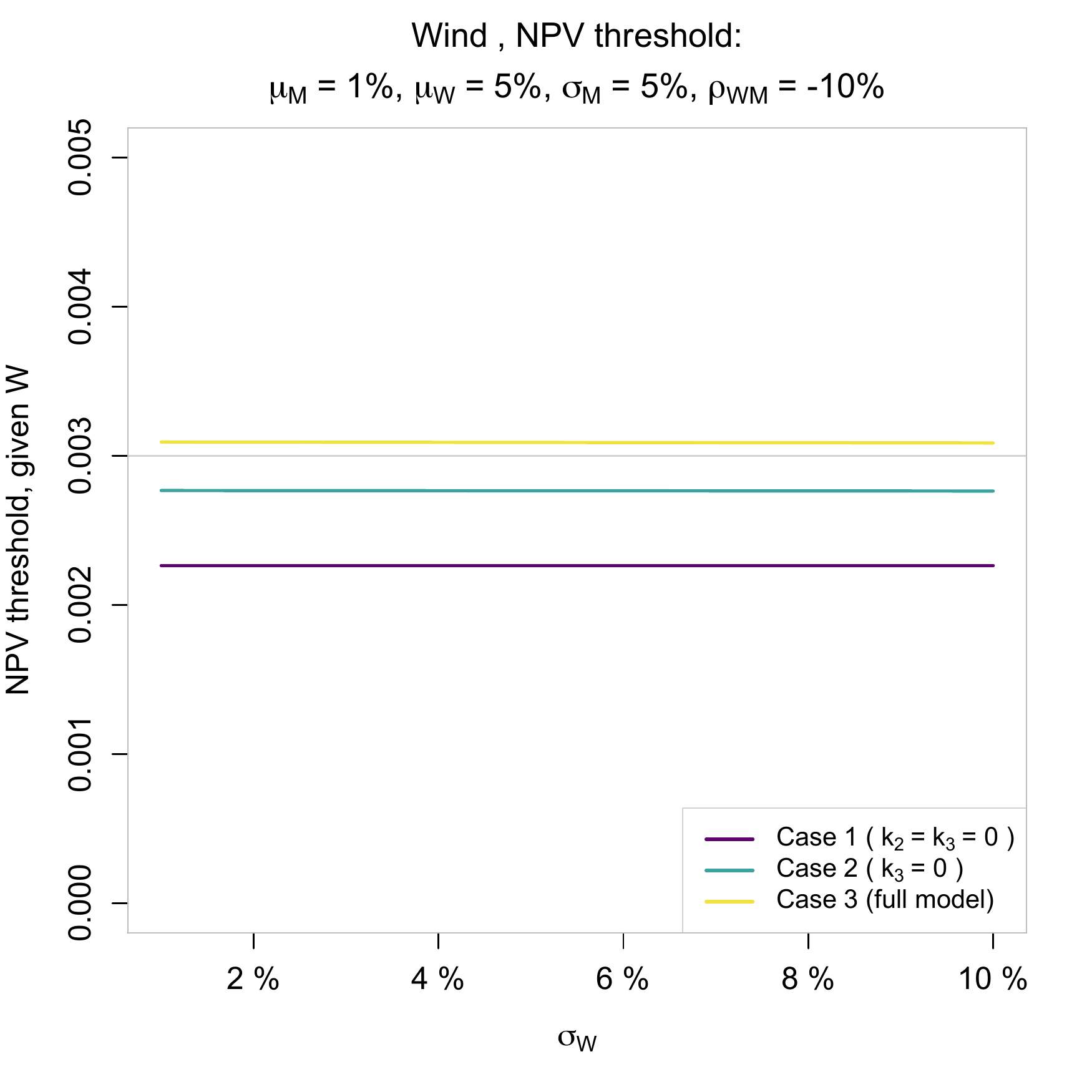}
\end{minipage}
\begin{minipage}{0.45\linewidth}
    \includegraphics[width=0.9\linewidth]{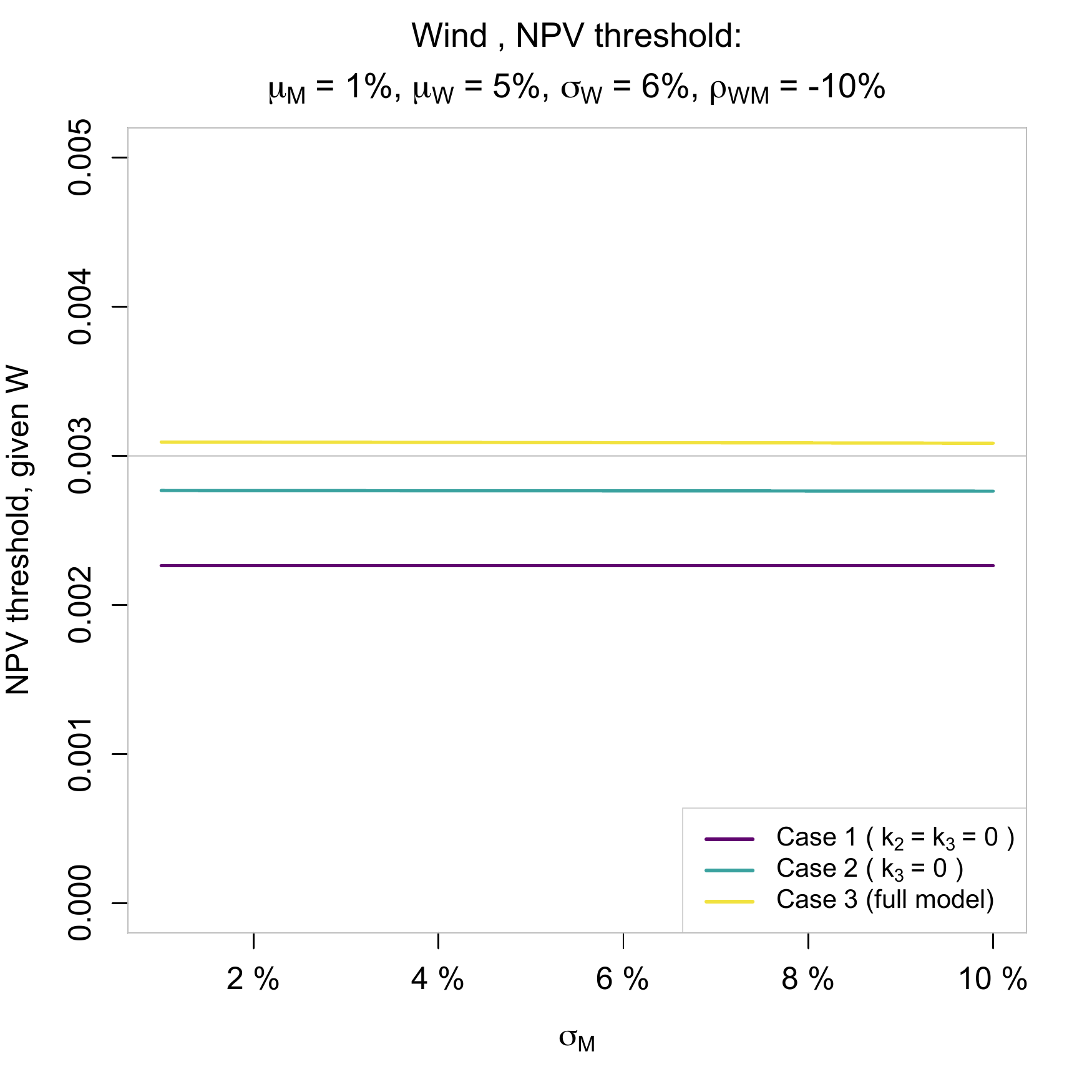}
\end{minipage}
\caption{Thresholds above which the NPV is positive. For the slope of the MO curve given VRE capacity level (in euros) for 1~MW wind power investment. Thresholds are presented for different values of $\mu_M$ and $\mu_W$ and standard deviations for $\sigma_M$ and $\sigma_W$ along the x-axes. Other parameters are given on top of each figure. Different colors indicate the full model (yellow), omitting cannibalization (teal), and omitting VRE merit-order effect and cannibalization (purple). The grey horizontal lines indicate the initial slope of the MO curve, $m_0$.}
\label{fig:Timing_npv}
\end{figure}

Investment thresholds according to the NPV rule are presented in \ref{fig:Timing_npv}.

\section{Model validation}
\label{sec:validation}

The primary issue with model validity is the emergence of negative prices when $G_{A,t} w_t \geq d$, as noted in Section \ref{sec:Short-term revenue}. While negative prices have become relatively frequent in markets with a high capacity of VRE \citep{Khoshrou2019}, stemming from e.g. inflexibility in demand and dispatchable generation and feed-in tariffs for VRE; it is not entirely obvious to which extent they should be present in our framework. With strict interpretation of our modeling framework and an assumption that VRE generation can be curtailed at will when prices go to zero \citep{EkholmVirasjoki2020}, price should be determined by $P_t = \max\{m_t (d - w_t G_{A,t}),0\}$. 

\begin{figure}[!htb]
\centering
    \includegraphics[width=0.9\linewidth]{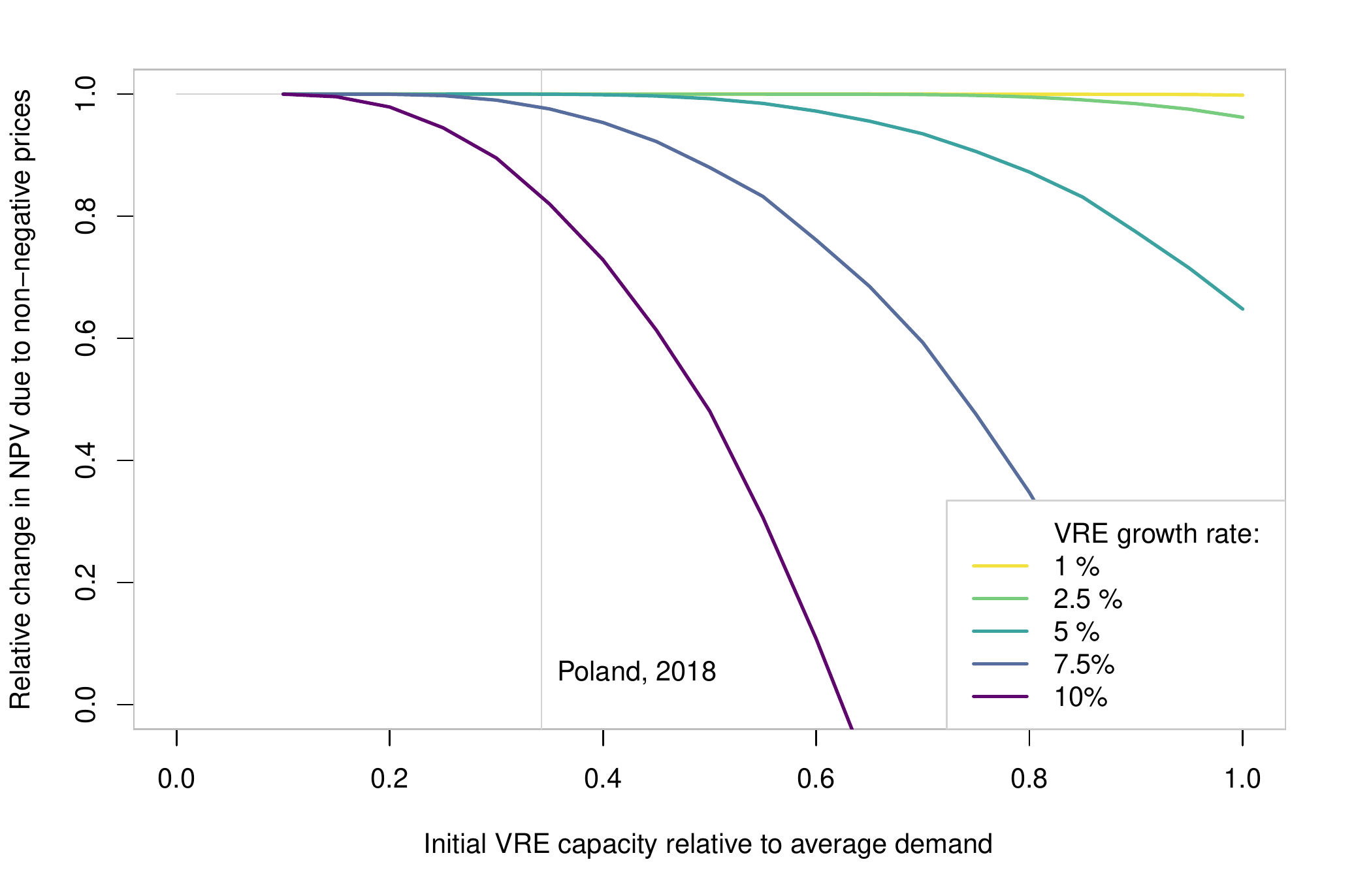}
\caption{Model error due to negative prices. The x-axis presents the initial VRE capacity $w_0$ relative to the average demand $d$. The y-axis presents the expected value from equation \eqref{eq:VRE_exp_revenue} relative to the corresponding result of Monte Carlo simulation that assumes non-negative prices. Colours indicate different rates of aggegate VRE capacity growth $\mu_W$.}
\label{fig:NPV_nonneg_error}
\end{figure}

To measure the different in the price formation -- i.e. the above formula compared to \eqref{eq:p_function_g} -- we ran a Monte Carlo simulation (sample size 1000, each with 2001 time steps over the VRE plant lifetime) that excludes negative prices as above, and compared thus calculated NPV of revenues to that given by equation \eqref{eq:VRE_exp_revenue}. The results are presented in Figure \ref{fig:NPV_nonneg_error}. As negative prices become problematic with only a high penetration of VRE, this experiment was done for a range of VRE capacity expansion rates $\mu_W$ (as in the numerical examples) and starting level of VRE capacity relative to the demand $m_0/d$. 

Given the initial capacity in the Polish case, VRE capacity expansion rates up to 7.5\% produce only minor differences due to negative prices, but with $\mu_W = 10\%$ the difference is already moderate, but still less than 20\%. Therefore one can conclude that the expected revenues in Figure \ref{fig:NPVRisk} towards $\mu_W = 10\%$ are slight underestimates if one deems that negative prices should be excluded from the model.

\clearpage
\section*{References}

\bibliographystyle{elsarticle-num}

\biboptions{sort&compress}
\bibliography{references}

\end{document}